\documentclass[12pt]{emulateapj}
\usepackage{graphics}
\usepackage{subfigure}
\usepackage{hyperref}
\usepackage{graphicx}
\usepackage{amsmath}
\usepackage{natbib}
\usepackage{float}
\usepackage{float}
\usepackage{amsmath}
\usepackage{natbib}
\usepackage{float}
\usepackage{gensymb}
\usepackage{color}

\begin{document}

\title{Integral Field Spectroscopy of the Low-mass Companion HD 984 B with the Gemini Planet Imager }

\author{Mara Johnson-Groh\altaffilmark{1}, Christian Marois\altaffilmark{2,1}, Robert J. De Rosa\altaffilmark{3},  Eric~L.~Nielsen\altaffilmark{4,5}, Julien~Rameau\altaffilmark{6}, Sarah Blunt\altaffilmark{7, 4, 5}, Jeffrey Vargas\altaffilmark{3}, S.~Mark~Ammons\altaffilmark{8}, Vanessa~P.~Bailey\altaffilmark{5}, Travis~S.~Barman\altaffilmark{9}, Joanna Bulger\altaffilmark{10}, Jeffrey~K.~Chilcote\altaffilmark{11},  Tara~Cotten\altaffilmark{12}, Ren\'{e}~Doyon\altaffilmark{6}, Gaspard Duch\^{e}ne\altaffilmark{3,13}, Michael~P.~Fitzgerald\altaffilmark{14}, Kate~B.~Follette\altaffilmark{5}, Stephen~Goodsell\altaffilmark{15,16},  James~R.~Graham\altaffilmark{3}, Alexandra~Z.~Greenbaum\altaffilmark{17}, Pascale~Hibon\altaffilmark{18}, Li-Wei~Hung\altaffilmark{14}, Patrick~Ingraham\altaffilmark{19}, Paul~Kalas\altaffilmark{3}, Quinn~M.~Konopacky\altaffilmark{20}, James~E.~Larkin\altaffilmark{14}, Bruce~Macintosh\altaffilmark{5}, J\'{e}r\^{o}me~Maire\altaffilmark{11}, Franck~Marchis\altaffilmark{4}, Mark~S.~Marley\altaffilmark{21}, Stanimir~Metchev\altaffilmark{22,23}, Maxwell A. Millar$-$Blanchaer\altaffilmark{24,11}, Rebecca~Oppenheimer\altaffilmark{25}, David~W.~Palmer\altaffilmark{8}, Jenny~Patience\altaffilmark{26}, Marshall~Perrin\altaffilmark{27}, Lisa~A.~Poyneer\altaffilmark{8}, Laurent Pueyo\altaffilmark{27}, Abhijith~Rajan\altaffilmark{26}, Fredrik~T.~Rantakyr\"o\altaffilmark{15}, Dmitry~Savransky\altaffilmark{28}, Adam~C.~Schneider\altaffilmark{26}, Anand~Sivaramakrishnan\altaffilmark{27}, Inseok~Song\altaffilmark{12}, Remi~Soummer\altaffilmark{27}, Sandrine~Thomas\altaffilmark{29}, David~Vega\altaffilmark{4}, J.~Kent~Wallace\altaffilmark{30}, Jason~J.~Wang\altaffilmark{3}, Kimberly~Ward-Duong\altaffilmark{26}, Sloane~J.~Wiktorowicz\altaffilmark{31} and Schuyler~G.~Wolff\altaffilmark{17}}

\affil{$^{1}$Department of Physics and Astronomy, University of Victoria, 3800 Finnerty Rd, Victoria, BC V8P 5C2, Canada}
\affil{$^{2}$Herzberg Astronomy \& Astrophysics, National Research Council of Canada, 5071 West Saanich Road., Victoria, BC V9E 2E7, Canada}
\affil{$^{3}$Department of Astronomy, UC Berkeley, Berkeley CA, 94720, USA}
\affil{$^{4}$SETI Institute, Carl Sagan Center, 189 Bernardo Avenue, Mountain View, CA 94043, USA}
\affil{$^{5}$Kavli Institute for Particle Astrophysics and Cosmology, Stanford University, Stanford, CA 94305, USA}
\affil{$^{6}$Institut de Recherche sur les Exoplan\`{e}tes, D\'{e}partment de Physique, Universit\'{e} de Montr\'{e}al, Montr\'{e}al QC H3C 3J7, Canada}
\affil{$^{7}$Department of Physics, Brown University, Providence, RI 02912, USA}
\affil{$^{8}$Lawrence Livermore National Laboratory, 7000 East Ave, Livermore, CA, 94550, USA}
\affil{$^{9}$Lunar and Planetary Lab, University of Arizona, Tucson, AZ 85721, USA}
\affil{$^{10}$Subaru Telescope, NAOJ, 650 North A'ohoku Place, Hilo, HI 96720, USA}
\affil{$^{11}$Dunlap Institute for Astronomy \& Astrophysics, University of Toronto, 50 St. George St, Toronto ON M5S 3H4, Canada}
\affil{$^{12}$Department of Physics and Astronomy, University of Georgia, Athens, GA}
\affil{$^{13}$Universit\'{e} Grenoble Alpes / CNRS, Institut de Plan\'{e}tologie et d'Astrophysique de Grenoble, 38000 Grenoble, France}
\affil{$^{14}$Department of Physics and Astronomy, UCLA, Los Angeles, CA 90095, USA}
\affil{$^{15}$Department of Physics, Durham University, Stockton Road, Durham DH1, UK} 
\affil{$^{16}$Gemini Observatory, Casilla 603, La Serena, Chile}
\affil{$^{17}$Physics and Astronomy Department, Johns Hopkins University, Baltimore MD, 21218, USA}
\affil{$^{18}$European Southern Observatory, Casilla 19001-Santiago 19-Chile}
\affil{$^{19}$Large Synoptic Survey Telescope, 950 N Cherry Ave, Tucson AZ 85719, USA}
\affil{$^{20}$Center for Astrophysics and Space Sciences, University of California, San Diego, La Jolla, CA 92093, USA}
\affil{$^{21}$Space Science Division, NASA Ames Research Center, Mail Stop 245-3, Moffett Field CA 94035, USA}
\affil{$^{22}$Department of Physics and Astronomy, Centre for Planetary Science and Exploration, The University of Western Ontario, London, ON N6A 3K7, Canada}
\affil{$^{23}$Department of Physics and Astronomy, Stony Brook University, Stony Brook, NY 11794-3800, USA}
\affil{$^{24}$Department of Astronomy \& Astrophysics, University of Toronto, Toronto ON M5S 3H4, Canada}
\affil{$^{25}$American Museum of Natural History, Department of Astrophysics, Central Park West at 79th Street, New York, NY 10024, USA}
\affil{$^{26}$School of Earth and Space Exploration, Arizona State University, PO Box 871404, Tempe, AZ 85287, USA}
\affil{$^{27}$Space Telescope Science Institute, 3700 San Martin Drive, Baltimore MD 21218 USA}
\affil{$^{28}$Sibley School of Mechanical and Aerospace Engineering, Cornell University, Ithaca, NY 14853}
\affil{$^{29}$Large Synoptic Survey Telescope, 950 N Cherry Ave, Tucson AZ, 85719, USA}
\affil{$^{30}$Jet Propulsion Laboratory, California Institute of Technology, 4800 Oak Grove Drive, Pasadena, CA 91109, USA}
\affil{$^{31}$The Aerospace Corporation, 2310 E. El Segundo Blvd., El Segundo, CA 90245, USA}




\begin{abstract}
We present new observations of the low-mass companion to HD 984 taken with the Gemini Planet Imager as a part of the Gemini Planet Imager Exoplanet Survey campaign.  Images of HD 984 B were obtained in the {\it J} (1.12--1.3 \micron) and {\it H} (1.50--1.80 \micron) bands.  Combined with archival epochs from 2012 and 2014, we fit the first orbit to the companion to find an 18~AU (70 year) orbit with a 68\% confidence interval between 14 and 28 AU, an eccentricity of 0.18 with a 68\% confidence interval between 0.05 and 0.47, and an inclination of 119\degree~with a 68\% confidence interval between 114\degree~and 125\degree.  To address considerable spectral covariance in both spectra, we present a method of splitting the spectra into low and high frequencies to analyze the spectral structure at different spatial frequencies with the proper spectral noise correlation.  Using the split spectra, we compare to known spectral types using field brown dwarf and low-mass star spectra and find a best fit match of a field gravity M$6.5\pm1.5$ spectral type with a corresponding temperature of $2730^{+120}_{-180}$ K.  Photometry of the companion yields a luminosity of $\log (L_{\mathrm{bol}}$/$L_{\sun}) = -2.88\pm0.07$~dex, using DUSTY models.  Mass estimates, again from DUSTY models, find an age-dependent mass of $34\pm1$ to $95\pm4$~M$_{\mathrm{Jup}}$.  These results are consistent with previous measurements of the object.
 
\end{abstract}

\section{Introduction}\label{intro}  

The search for exoplanets through direct imaging has led to many serendipitous detections of brown dwarfs and low-mass stellar companions \citep[e.g.][]{chauvin05,biller,nielsen,mawet,konopacky16}.  These surveys tend to target young, bright stars whose potential companions would still be warm and bright in the infrared \citep[e.g.,][]{burrows97}.  Brown dwarfs, having higher temperatures than planetary-mass companions of the same age, are significantly brighter and therefore easier to detect.  While brown dwarf companions are not the primary focus of direct imaging searches, they are useful in their own right for a better understanding of substellar atmospheres and for comparing competing formation models \citep[e.g.][]{perets,bodenheimer,boley}.

\citet{meshkat} reports the discovery of a bound low-mass companion to HD 984, a bright, nearby ($47.1\pm1.4$ pc) F7V star of mass $\sim$1.2 M$_\sun$ \citep{van,meshkat} and a temperature of $6315\pm89$~K \citep{white,casagrande}.  With an age estimate of 30--200 Myr ($115\pm85$~Myr at a 95\% confidence level) derived from isochronal age, X-ray emission and rotation \citep{meshkat}, and consistent with previous age estimates \citep{wright04,torres}, HD 984 is ideal for direct imaging campaigns to search for young substellar objects.

The results presented in \citet{meshkat} finds HD 984 B at a separation of $0\farcs19\pm0\farcs02$ ($9.0\pm1.0$~au) based on the $L'$ observations with NaCo \citep{lenzen,rousset} in 2012 July, and the $H+K$ band observations with SINFONI  \citep{eisenhauer,bonnet} in 2014 September, both mounted on UT4 on the Very Large Telescope (VLT) at Cerro Paranal, Chile. Comparing the SINFONI spectrum to field brown dwarfs and low-mass star in the NASA Infrared Telescope Facility (IRTF) library, \citet{meshkat} conclude the companion to be a M$6.0\pm0.5$ object \citep{cushing,rayner}. The paper reports an estimated mass of $33-96$ M$_\mathrm{Jup}$, overlapping the brown dwarf and low-mass star regime. This mass range corresponds to a mass-ratio of $q=0.03-0.08$. \citet{meshkat} note that future observations in the $J$ band could provide additional constraints on the surface gravity of the companion, which could help place the system in the lower or higher range of the age estimate of $115\pm85$~Myr.

Competing theories of companion formation are reliant upon observational studies to verify their models. Systems like HD 984 B, with low mass-ratios are particularly interesting for understanding brown dwarf and low-mass companion formation. \citet{kraus} find a flat mass-ratio distribution for solar type stars, which they state to be consistent with formation via fragmentation on small scales, however their sample does not extend below $q=0.1$. Furthermore, the long baseline of observations of HD 984 B allows for better determination of the companion's orbit. This is necessary for understanding formation scenarios, given their dependence on separation, as some suggest disk instability is responsible for companion formation at large ($\gtrapprox50 $ au) separations and protostellar core fragmentation at small ones ($\lessapprox50$ au) \citep[e.g.][]{boss,bod, clarke}. Finally, characterizing young binary star systems like HD 984 is important for future observations with JWST and$/$or 30-m class telescopes that could discoverer lower mass exoplanets, giving insight of planet formation in binary systems.

HD 984 was observed as one of the targets of the Gemini Planet Imager Exoplanet Survey (GPIES, \citealp{macintosh14}), an ongoing survey of over 600 nearby young stars with nearly 900 hours dedicated time on the 8-meter Gemini South Telescope using the Gemini Planet Imager \citep[GPI,][]{macintosh}, a coronagraphic adaptive optics integral field spectrograph and imaging polarimeter. Using near-infrared imaging spectroscopy (0.9--2.4 \micron) and advanced imaging and post-processing techniques, GPI detects thermal emission from exoplanets and brown dwarfs at angular separations of $0\farcs2$--$1\farcs0$ from their parent star.

We present new spectroscopic observations of HD 984 B at $J$ (1.12--1.35 \micron, $R\sim34$) and $H$ (1.50--1.80 \micron, $R\sim45$) bands with GPI, from which we derive the first orbit estimate and update luminosity, magnitude and mass measurements.  In Section~\ref{obsv} we describe the GPI observations. Basic reductions are explained in Section~\ref{redu}.  Section~\ref{psfsub} details the point spread function (PSF) subtraction technique.  Astrometry is discussed in Section~\ref{astrometry}, orbital fitting in Section~\ref{orbitp} and spectral and photometric analyses are presented in Section~\ref{analysis}. Finally, we conclude in Section~\ref{conc}.


\section{Observations}\label{obsv}
HD 984 was observed with the GPI integral field spectrograph \citep[IFS,][]{chilcote,larkin} on 2015 August 30 UT during the GPIES campaign (program GPIES-2015B-01, Gemini observation ID: GS-2015B-Q-500-982) at Gemini South.  The GPI IFS has a field-of-view (FOV) of $2.8\times2.8$ arcsec$^2$ with a plate scale of $14.166\pm0.007$ miliarcseconds/pixel and a position angle offset of $-0\fdg10\pm 0\fdg13$ \citep{macintosh, derosa15}.  Coronographic images were taken in spectral mode in the $J$ and $H$ bands.  Observations were performed when the star was close to the meridian at an average airmass of 1.1 so as to maximize FOV rotation for angular differential imaging \citep[ADI,][]{marois06} and minimize the airmass during observations.  Twenty-three exposures of 60 seconds of one coadd each were taken in the {\it H} band and 23 exposures, also of 60s and one coadd, were followed up in the $J$ band; the $J$ band data was acquired with the $H$ band apodizer due an apodizer wheel mechanical issue. Two $H$ band exposures and five $J$ band exposures were rejected due to unusable data quality.  Total FOV rotation for {\it H} band was $15\fdg2$ and a total rotation of $11\fdg9$ was acquired with {\it J} band. Average DIMM seeing for $H$ and $J$ band sequences was $1\farcs14$ and $0\farcs82$ respectively, higher than the median seeing of $0\farcs65$. The windspeed averages for $H$ and $J$ bands were 2.5 m s$^{-1}$ and 1.9 m s$^{-1}$.  Images for wavelength calibration were taken during the daytime at zenith and short exposure arc images were acquired just before the sequences at the target elevation to correct for instrument flexure \citep{wolff}.


\section{Reductions}\label{redu}
The images were reduced using the GPI Data Reduction Pipeline \citep{perrin} v1.3.0\footnote{\href{http://docs.planetimager.org/pipeline}{\url{http://docs.planetimager.org/pipeline}}}.  Using primitives in the pipeline, raw images were dark subtracted, argon arc image comparisons were used to compensate for instrument flexure \citep{wolff}, the spectral data cube was extracted from the 2D images \citep{maire}, bad pixels were interpolated in the cube and distortion corrections were applied \citep{konopacky}.   A wavelength solution was obtained using arc lamp images taken during the day prior to data acquisition. Four satellite spots, PSF replicas of the star generated by the pupil-plane diffraction grating \citep{marois06,siv},  were used to measure the location of the star behind the coronagraphic mask for image registration at a common centre, and to calibrate the object flux to star flux \citep{wang}.



\section{PSF Subtraction}\label{psfsub}
\begin{figure*}
  \centering
  \begin{tabular}[b]{@{}p{0.35\textwidth}@{}}
   \includegraphics[width=1.0\linewidth]{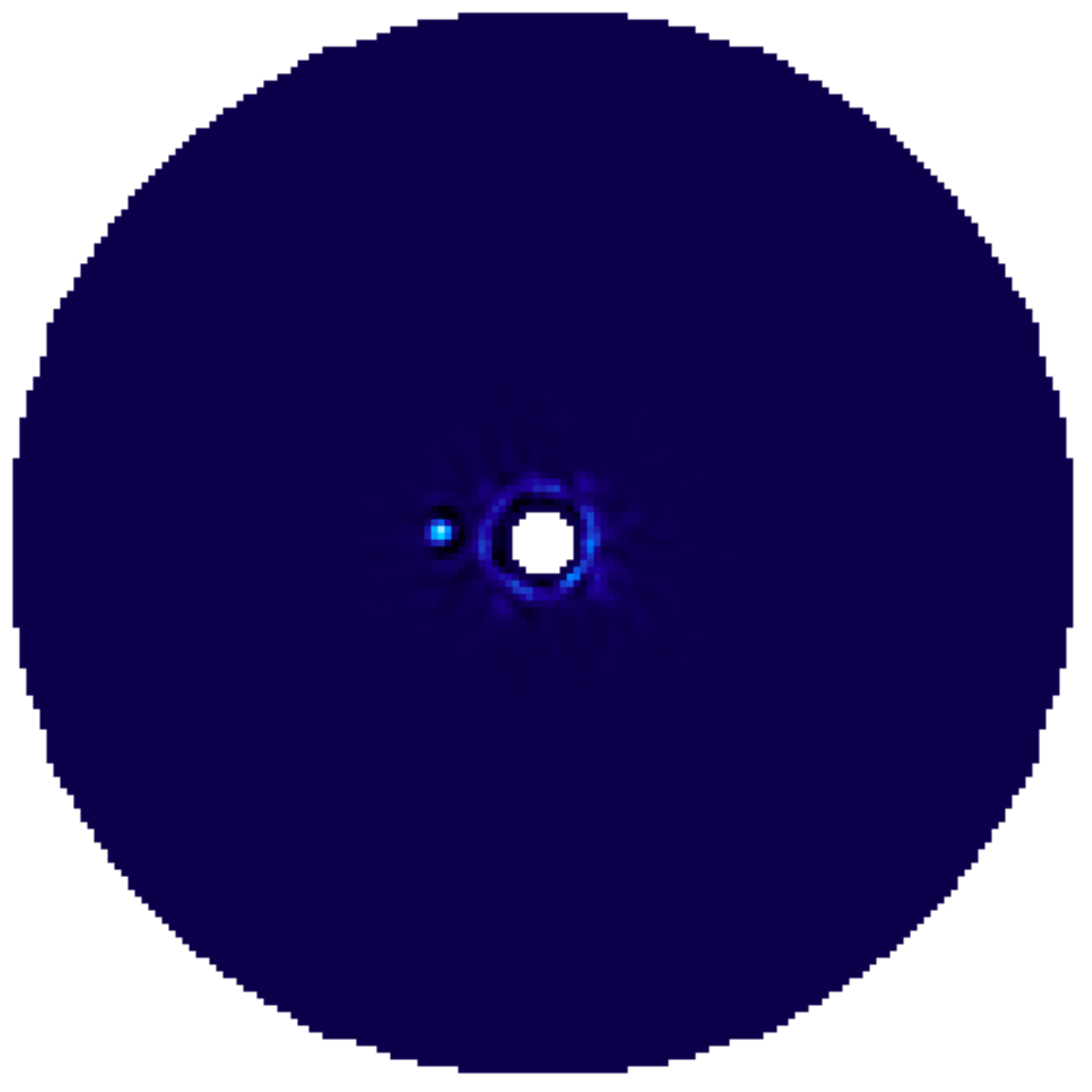} \\
    \centering\small (a)  
  \end{tabular}%
  \quad
  \begin{tabular}[b]{@{}p{0.35\textwidth}@{}}
    \includegraphics[width=1.0\linewidth]{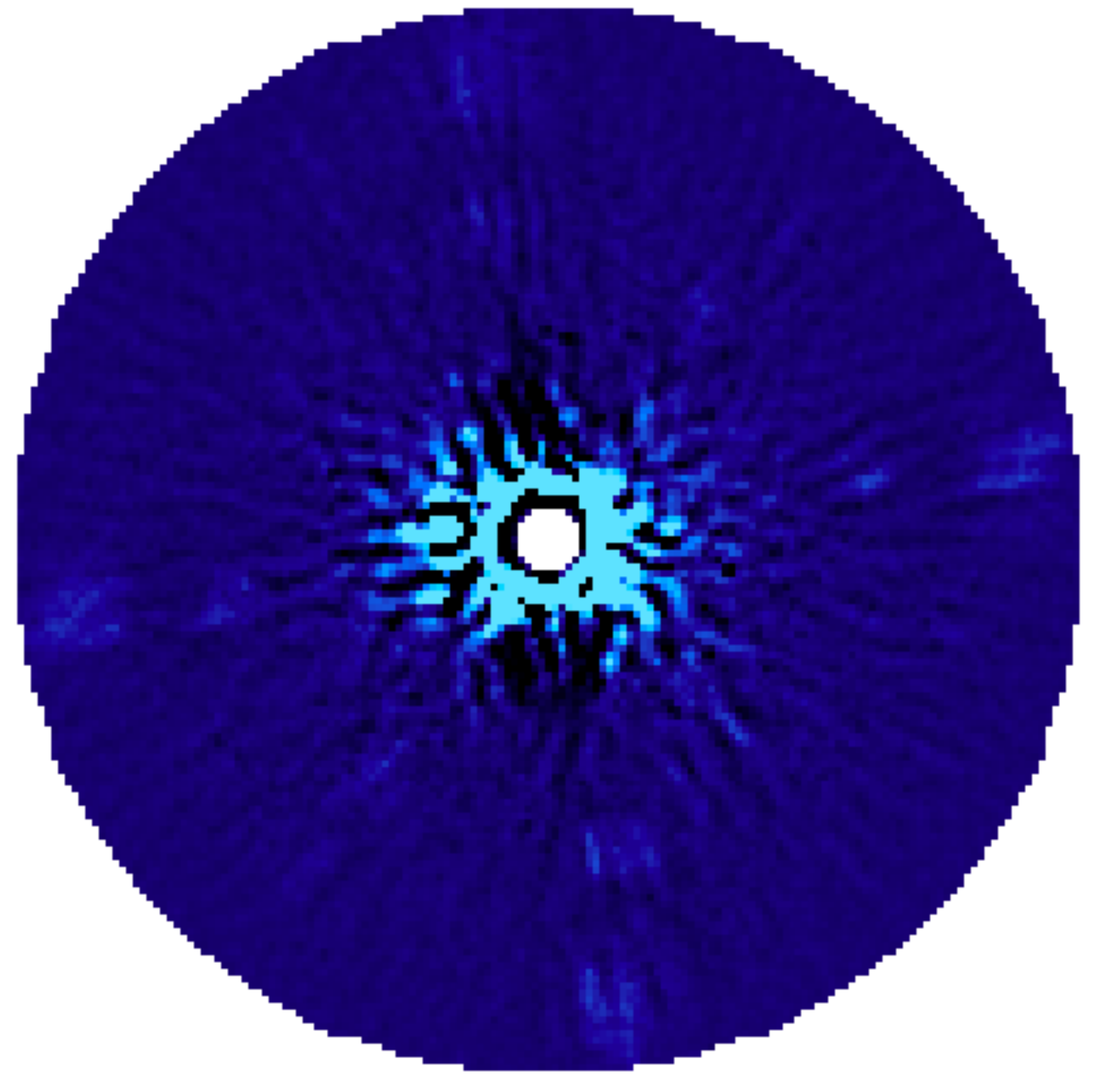} \\
    \centering\small (b) 
  \end{tabular} \\
    \begin{tabular}[b]{@{}p{0.35\textwidth}@{}}
   \includegraphics[width=1.0\linewidth]{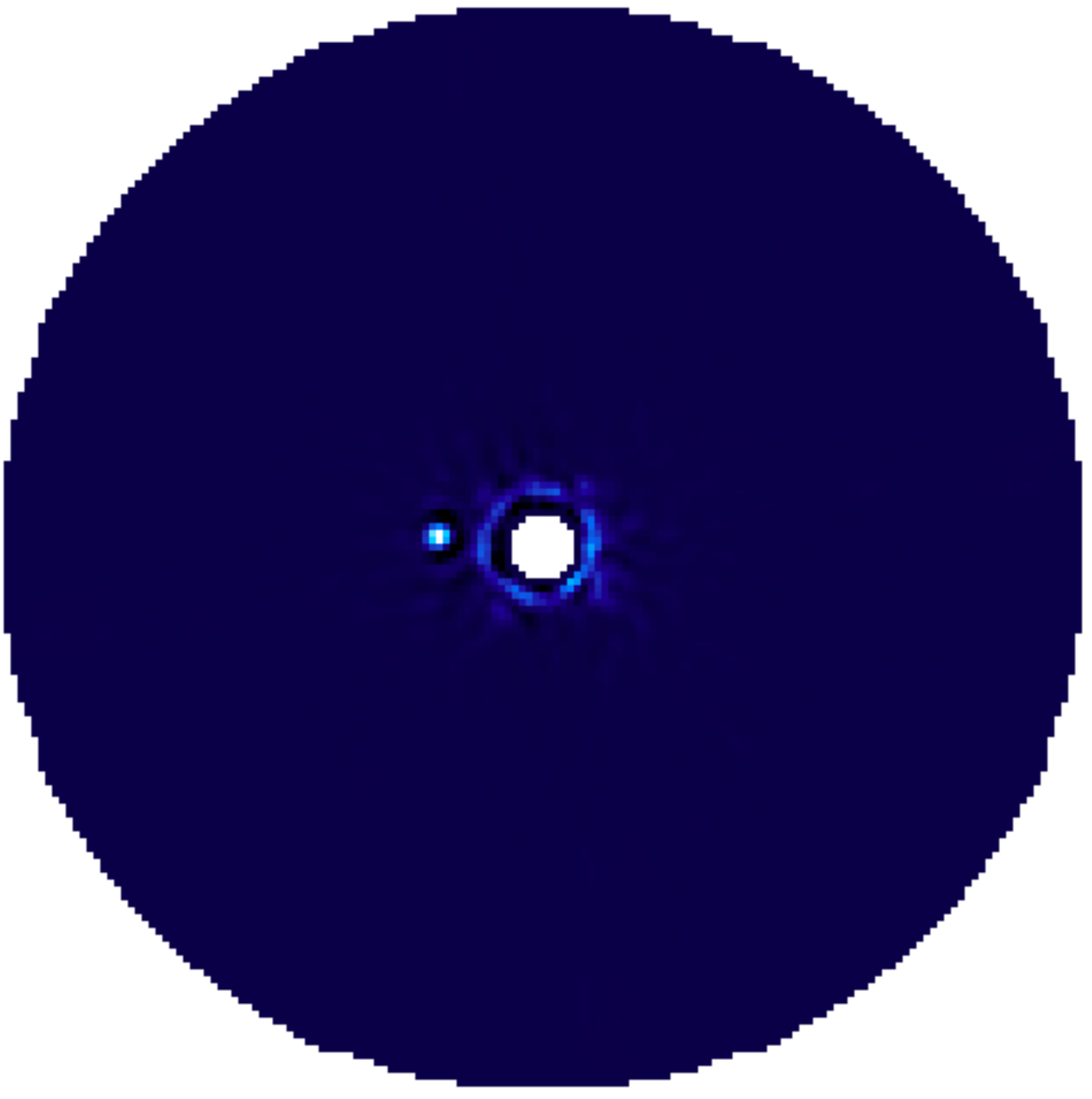} \\
    \centering\small (c)  
  \end{tabular}%
  \quad
  \begin{tabular}[b]{@{}p{0.35\textwidth}@{}}
    \includegraphics[width=1.0\linewidth]{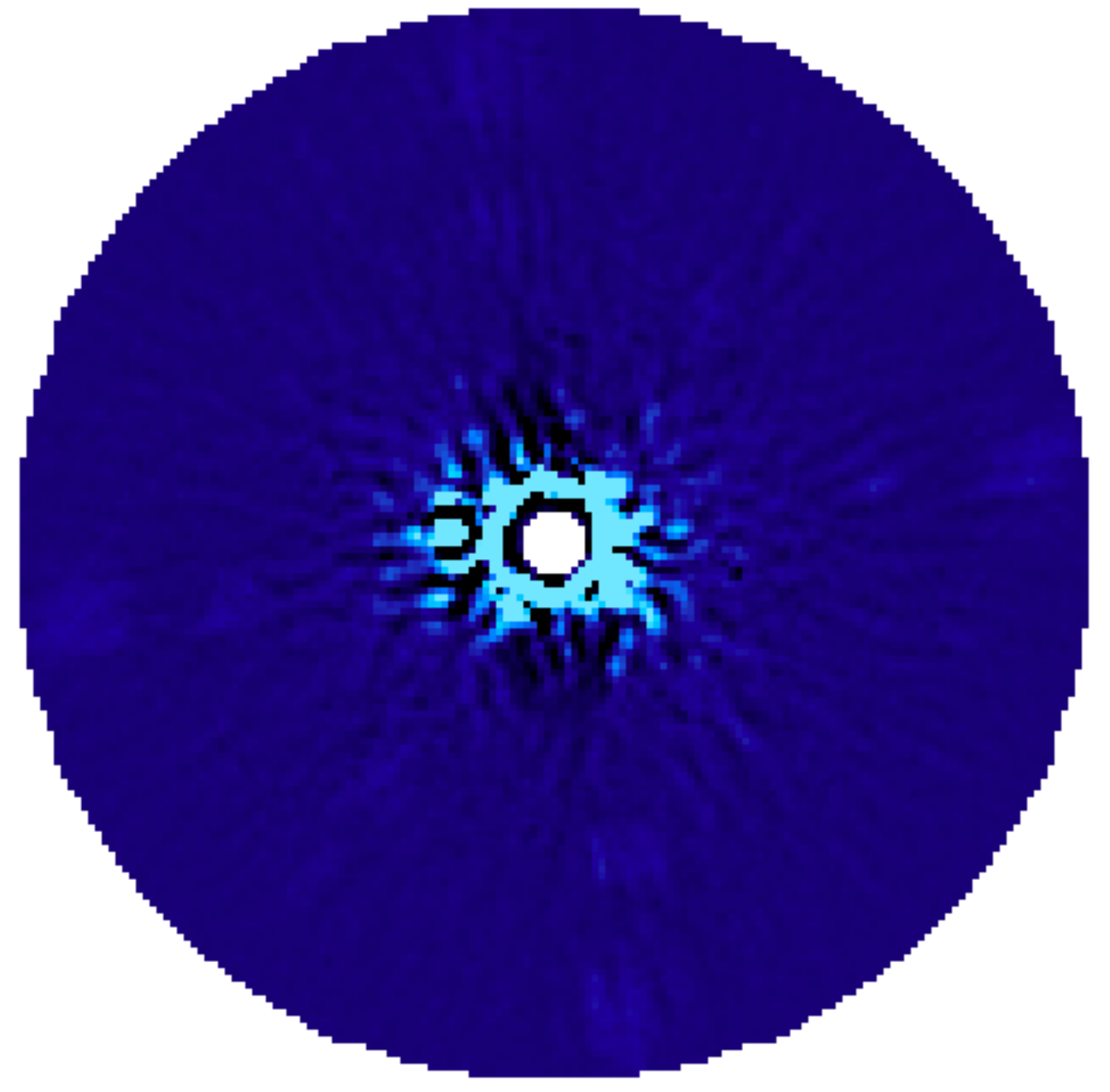} \\
    \centering\small (d) 
  \end{tabular}
  \caption[HD 984 B Final Images]{HD 984 B GPI $H$ band (a \& b) $J$ band (c \& d) final 2D images, post TLOCI using only an ADI subtraction. Two contrast levels are shown to highlight the companion (left) and the image noise (right). North is up and east is left. Each image has a diameter of $2\farcs23$.}
  \label{finalimages}
\end{figure*}

After the initial data reduction, each slice of each data cube, which were each flux normalized using the average maximum of a Gaussian fit on the four calibration spots, were spatially magnified to align diffraction-induced speckles using the pipeline-derived spot positions to determine the star position.  The images were then unsharp masked using a $11\times11$ pixel kernel to remove the seeing halo and background flux, and were PSF subtracted using the TLOCI (Template Locally Optimized Combination of Images) algorithm \citep{marois14}.  TLOCI uses \textit{a priori} planet spectral information to optimize the least squares-based subtraction algorithm to maximize planet detection when using data acquired with the simultaneous spectral differential imaging \citep[SSDI,][]{racine,marois00} and the ADI \citep{marois06} techniques. T8 and L0 templates created from the Brown Dwarf Spectroscopic Survey \citep{mclean} were used for these reductions.  The TLOCI pipeline was run in a 6 pixel (85 mas) width annulus of increasing size with an inner gap to avoid the coronagraph focal plane mask. The IDL \texttt{invert.pro} algorithm, which uses a Gaussian elimination method, was used to invert the correlation matrix with the single value decomposition cutoff algorithm which shows limited gains in the signal-to-noise ratio (SNR) for GPI data. Two annuli, one with a width of 5 pixels interior to the subtraction zone inner annuli, and another of a 10 pixel width exterior to the subtraction zone outer annuli, are used for the reference pixels to derive the correlation matrix (pixels in the subtraction zones are not included in the correlation matrix to avoid the algorithm to fit the planet).  Once all the data cubes had been PSF subtracted, a final 2D image was obtained by performing a weighted-mean of the 37 slices, using the input template spectrum and image noise to maximize the object's SNR. Final images for $J$ and $H$ bands shown in Figure~\ref{finalimages}. While the initial discovery was obtained by performing both an SSDI and ADI subtractions, to avoid spectral cross-talk bias, only a less aggressive ADI-only subtraction (reference images are selected if they have less than 30\% of the substellar object flux in a 1.5 $\lambda/D$ diameter aperture centred at the object position) was used for spectral and astrometry extractions. 

\section{Astrometry and Spectral Extraction}\label{astrometry}
Using the TLOCI ADI-only subtracted final combined data cube, the flux and position of the companion were measured relative to the star. As TLOCI uses a training zone that differs from the subtraction zone, the companion is not fitted by the least-squares, thus removing one bias. To further take into account the self-subtraction bias for both spectral and astrometry extraction, the companion's signal was fitted using a forward model derived from the median-average PSF of the four calibration spots for the entire sequence. For each slice of each data cube, a noiseless image was created with the calibration spot PSF at the approximate location of the companion. These simulated images were then processed using the same steps as the science images to produce the companion forward model. The forward model was then iterated in flux and position to minimize the local residual post subtraction in a 1.5 $\lambda/D$ circular aperture, where $\lambda$ is the wavelength and $D$ is the telescope diameter. Error bars were derived by adding and extracting the forward model flux at the same separation as the companion, but at nine different position angles.  The standard deviation in flux and position of the nine simulated companions are adopted as the spectral and astrometric errors. A correction is applied to account for forward model errors. Instead of adding the simulated companions having the same flux as the recovered flux, they are added after normalizing the companion signal by the ratio of the local residual noise inside a 1.5 $\lambda/D$ aperture after the forward model best subtraction relative to the noise at the same angular separation calculated away from the companion.  

In the $H$ band, the separation was $216.3\pm1.0$ mas, and in the $J$ band, $217.9\pm0.7$ mas.  The position angles for $H$ and $J$ bands were $83\fdg3\pm0\fdg3$ and $83\fdg6\pm0\fdg2$, respectively.  The uncertainty is a combination of measurement error added in quadrature with plate scale, north angle error, and star position error \citep[0.05 pixels,][]{wang}.  The astrometry is summarized alongside the previous measurements from \citet{meshkat} in Table~\ref{astrom}.
\begin{deluxetable*}{cccccl}
\tabletypesize{\scriptsize}
\tablewidth{0pt}
\setlength{\tabcolsep}{3pt}
\tablecaption{Astrometry}
\tablehead{{Band} &{Date} &{Separation (mas)}&{PA($\degree$)}&{Instrument}&{Note}}
\startdata
\hline
$L'$	&	2012 July 18    &	$190\pm20$	    &$108.8\pm3.0$  &   NaCo 	&APP data \\
$L'$	&	2012 July 20	&	$208\pm23$	    &$108.9\pm3.1$  &   NaCo 	&direct imaging \\
$HK$   &   2014 Sep 9		&   $201.6\pm0.4$	&$92.2\pm0.5$   &   SINFONI\\
$H$	    &	2015 Aug 29 	&	$216.3\pm1.0$	&$83.3\pm0.3$   &   GPI  \\
$J$	    &	2015 Aug 29	    &	$217.9\pm0.7$	&$83.6\pm0.2$   &   GPI \\
\enddata
\label{astrom}
\tablecomments{Data from 2012 and 2014 epochs from \citet{meshkat}.}
\end{deluxetable*}


 Since TLOCI does not directly yield the spectrum of the companion, but rather the ratio of the companion to the spectrum of the primary star, the stellar spectra therefore needs to be divided out.This calibration uses the known stellar spectral shape to recover the true shape of the object's spectra and was done using a custom IDL program and the Pickles stellar spectral flux library \citep{pickles}. First, spectra for both F5V and F8V models in $J$ and $H$ bands were degraded to GPI spectral resolution and interpolated at the same wavelengths as the GPI wavelength channels. 
The zero points were computed with a Vega spectrum\footnote{\url{http://www.stsci.edu/hst/observatory/crds/calspec.html}} and the $J$ and $H$ filter transmission curves and used to calibrate the Pickles template to the $J$ and $H$ magnitudes of HD\,984. For each band, the normalized F5V and F8V spectra were interpolated to compute a F7V since the Pickles library did not contain a model spectrum for an F7V star.  Finally, the calibrated F7V model spectrum was multiplied by the planet-to-star spectra for each band extracted from the GPI data to obtain the final {\it J} and {\it H} band spectra of HD~984~B.

\section{Orbital Fitting}\label{orbitp}
We determined the orbital parameters consistent with the full astrometric record of HD~984 using the rejection sampling method previously presented in \citet{derosa16} and \citet{rameau2016}. Orbital parameters were drawn from priors that are uniform for eccentricity ($e$), argument of periastron ($\omega$), and epoch of periastron passage ($T_0$), and uniform in $\cos i$ for inclination angle.  Semimajor axis ($a$) and position angle of nodes ($\Omega$) are assigned initial values of $a = 1$~au and $\Omega = 0$~degrees. These values are then adjusted so that the orbit matches one of the epochs of data which has the effect of imposing a prior uniform in position angle of nodes and uniform in $\log a$.   Observational errors are taken into account by scaling to Gaussian distributions in separation and position angle, centered on the measurements from the reference epoch with the standard deviations of the Gaussians equal to the errors.  Period is not fit independently but is instead derived from Kepler's third law and the mass of the star of 1.2 $M_\sun$.

\begin{figure}[h]
\epsscale{1.2}
\plotone{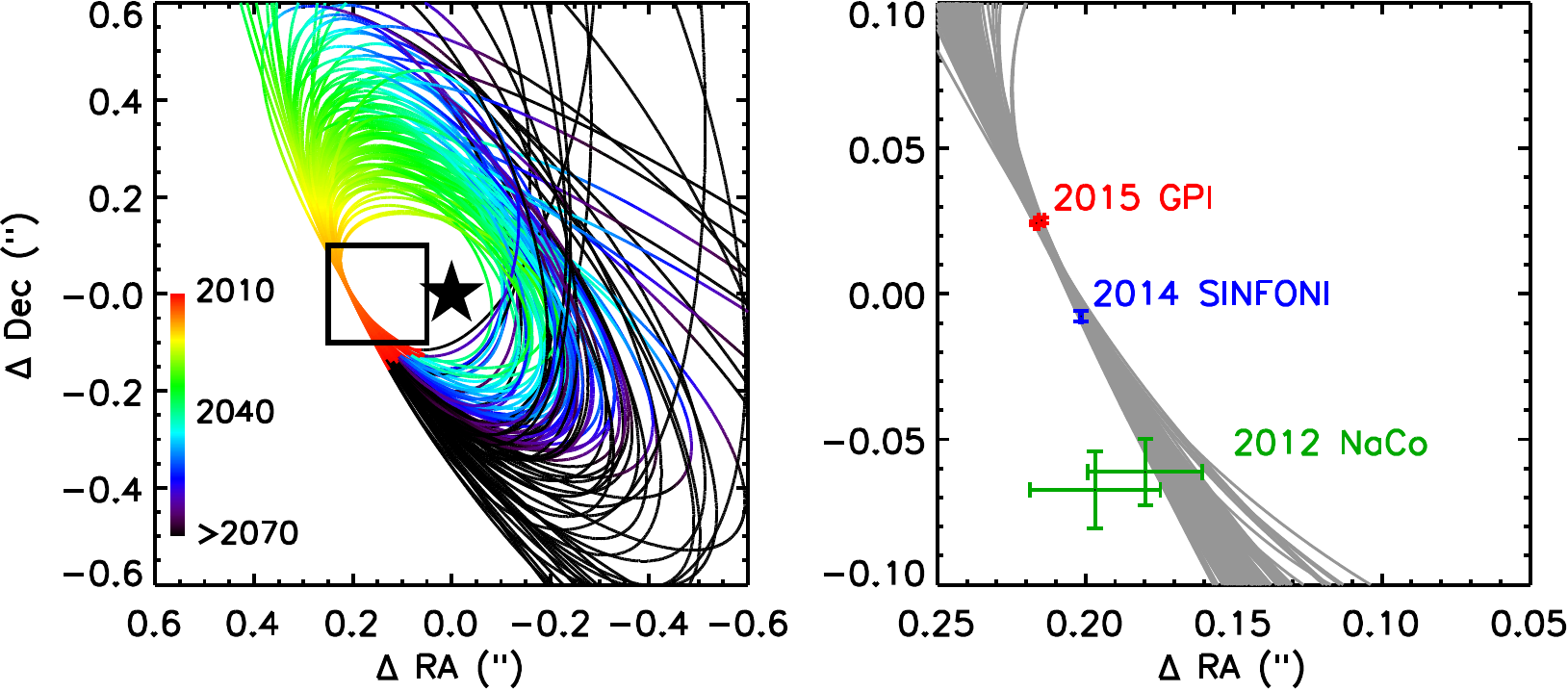}
\caption[Orbital Models for HD 984 B]{Orbits drawn from the posteriors fit to the NaCo, SINFONI and GPI epochs.  Color in the left panel corresponds to the epoch that the companion reaches a given location.  The small square box in the left panel shows the range of the panel to the right. The resulting fit has a median 18~au (70 year) orbit, with a 68\% confidence interval between 14 and 28 au, an eccentricity of 0.18 with a 68\% confidence interval between 0.05 and 0.47, and an inclination of 119\degree~with a 68\% confidence interval between 114\degree~and 125\degree.   }
\label{orbit}
\end{figure}

\begin{figure}[h]
\epsscale{1}
\plotone{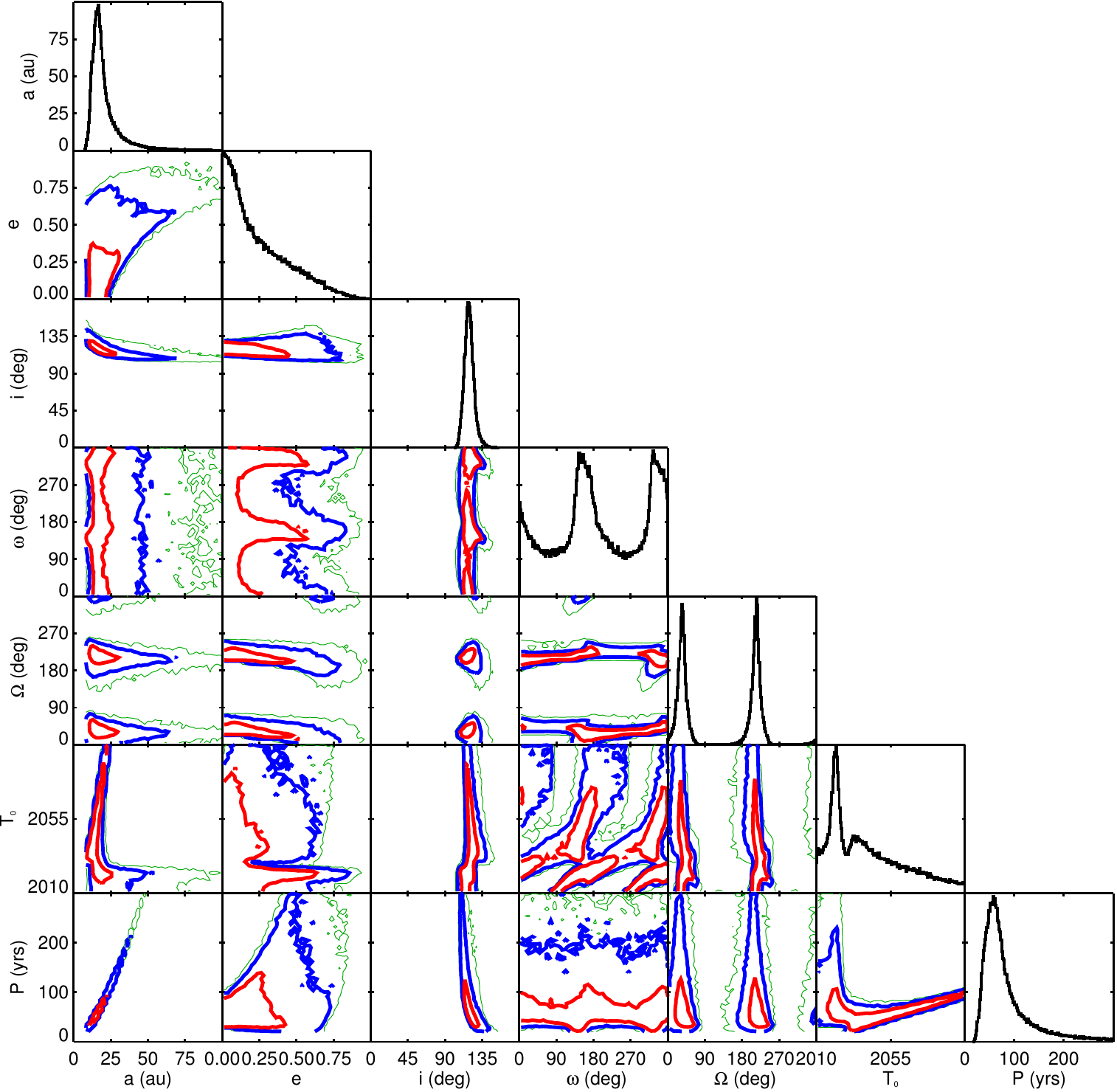}
\caption[Orbital Fitting Confidence Intervals]{Posterior probability distribution for the six orbital parameters fit by our rejection sampling method, and period derived using Kepler's third law and the mass of the star of 1.2 $M_\sun$.  For the off-diagonal panels, 1 (red), 2 (blue), and 3 (green) $\sigma$ contours enclose 68.27\%, 95.45\%, and 99.70\% of all orbital elements. }
\label{intervals}
\end{figure}

The fitting gives a median semimajor axis of 18 au (70 year) orbit, with a 68\% confidence interval between 14 and 28 au and an eccentricity of 0.18 with a 68\% confidence interval between 0.05 and 0.47 and inclination of 119$^\circ$ with a 68\% confidence interval between 114$^\circ$ and 125$^\circ$.  The highest probability orbit has parameters of $a=23.17$ au, $e=1.74$, $i=115\fdg26$, $\omega=49\fdg92$, $\Omega=208\fdg48$, $T_0=2105.80$, $P=101.80$ yrs, with $\chi^2=9.80$.  Fits to the orbit are shown in Figure~\ref{orbit} and posterior probability distributions and covariances of the orbital parameters are shown in Figure~\ref{intervals}.  \citet{meshkat} do not perform an orbital fit in their analysis, but from their two epochs believe the system to have a non-zero inclination, which we have confirmed here. 



\section{Photometric and Spectroscopic Analysis}\label{analysis}

\begin{figure}[h]
\epsscale{1.2}
\plotone{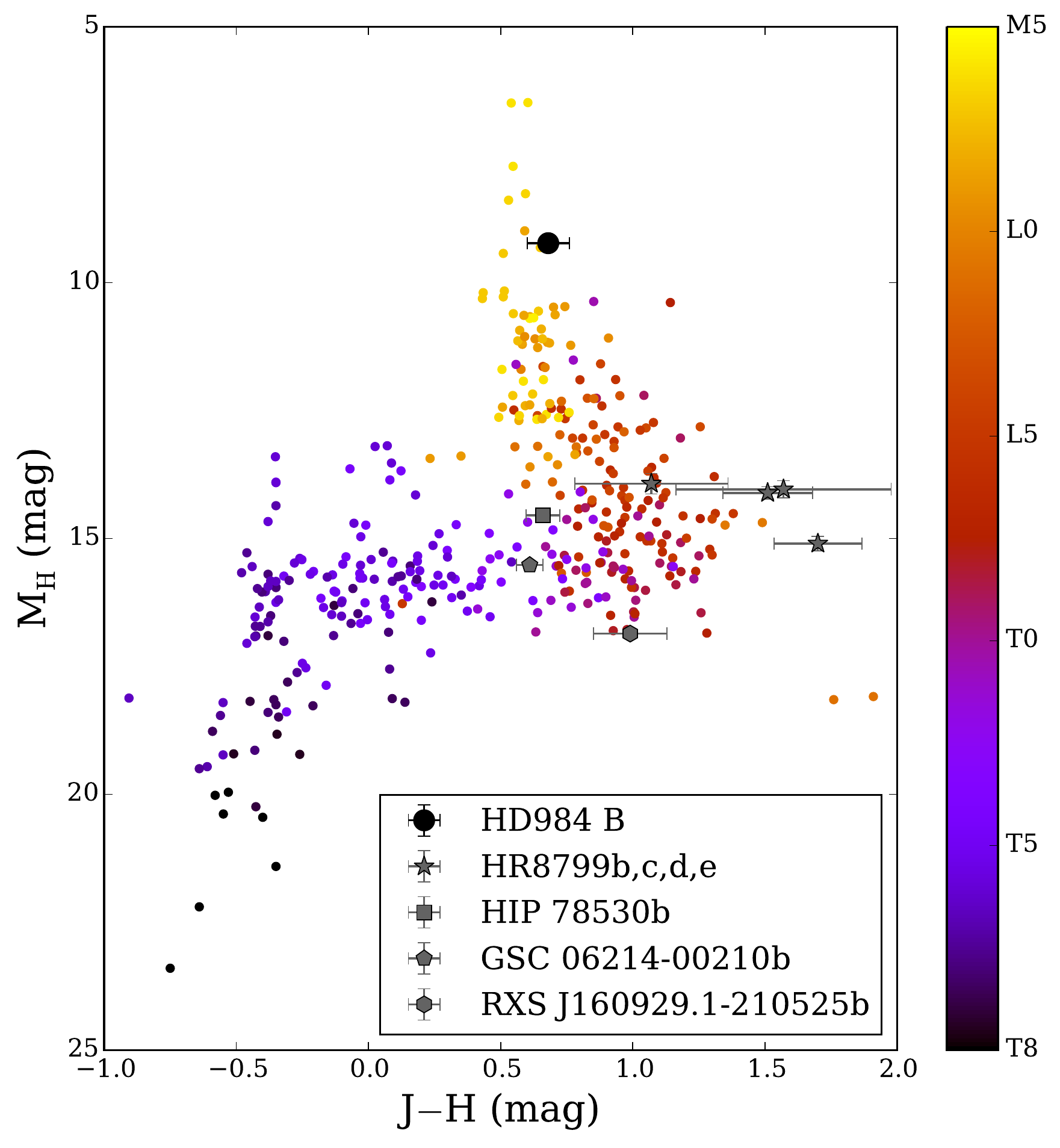}
\caption[CMD for HD 984 B]{$J-H$ colour magnitude diagram showing HD 984 B relative to other known brown dwarfs and low mass stars \citep{dupuy}.  Brown dwarf and low mass star spectral types are colour coded on a spectrum from dark purple (T types) to yellow (M types).  HD 984 B is shown as a black circle and is located on the late M/early L dwarf cooling sequence.  Photometry for other planets is from \citet{zurlo} (HR 8799 b,c,d,e), and \citet{lachapelle} (HIP 78530 b, GSC 06214-00210 b, RXS J160929.1-210525 b).}
\label{cmd}
\end{figure}

We first calculate the absolute magnitudes for HD~984~B by integrating the companion-to-star spectra, and correcting for the GPI filter transmission profile and Vega zero points \citep{derosa16}.  The $J$ and $H$ band apparent magnitudes were calculated to be $13.28\pm0.06$ and $12.60\pm0.05$, respectively.  Assuming a distance to the star of $47.1\pm1.4$ pc \citep{van}, the absolute magnitudes of the object in $J$ and $H$ bands are $9.92\pm0.09$ and $9.23\pm0.08$, respectively.  The $H$ band magnitude is consistent with the $H$ band magnitude reported in \citet{meshkat}.  A rudimentary spectral type can be ascertained using the $J$ and $H$ band magnitudes as compared to other brown dwarfs and low mass stars via a colour-magnitude diagram (see Figure~\ref{cmd}).  When compared with literature brown dwarfs and low-mass star from \citet{dupuy}, these magnitudes further corroborate the \citet{meshkat} result of a late M-type object.

For further characterization of the companion, we discuss the spectral analysis here.  A more detailed analysis of the spectral type requires attention to spectral noise covariance, which arises from the coupling of neighbouring wavelength channels in the spectra and is a result of the finite resolution of GPI.  This type of correlation correction is necessary for proper error calculation. IFS instruments observations often produce spectral noise covariance \citep[e.g.][]{greco16}, and GPI data cubes are also known to suffer from this effect, especially at small separations close to the focal plane mask.  Before any comparisons to field objects can be made, this spectral noise covariance needs to be characterized to avoid biasing any analysis with improper error calculations. Given the high SNR of detections, it may be possible to fit higher frequency structures (e.g. spectral lines) in the spectrum independently to the low frequency envelope (the overall shape of the spectra). The noise characteristics, especially the spectral noise correlation, may differ with spectral frequencies, with the low frequencies mostly limited by highly correlated speckle noise slowly moving over the object as a function of wavelength, while higher frequencies could be mainly limited by read or background noises, thus being weakly correlated between wavelength channels.

\begin{figure}[h]
\epsscale{1}
\plotone{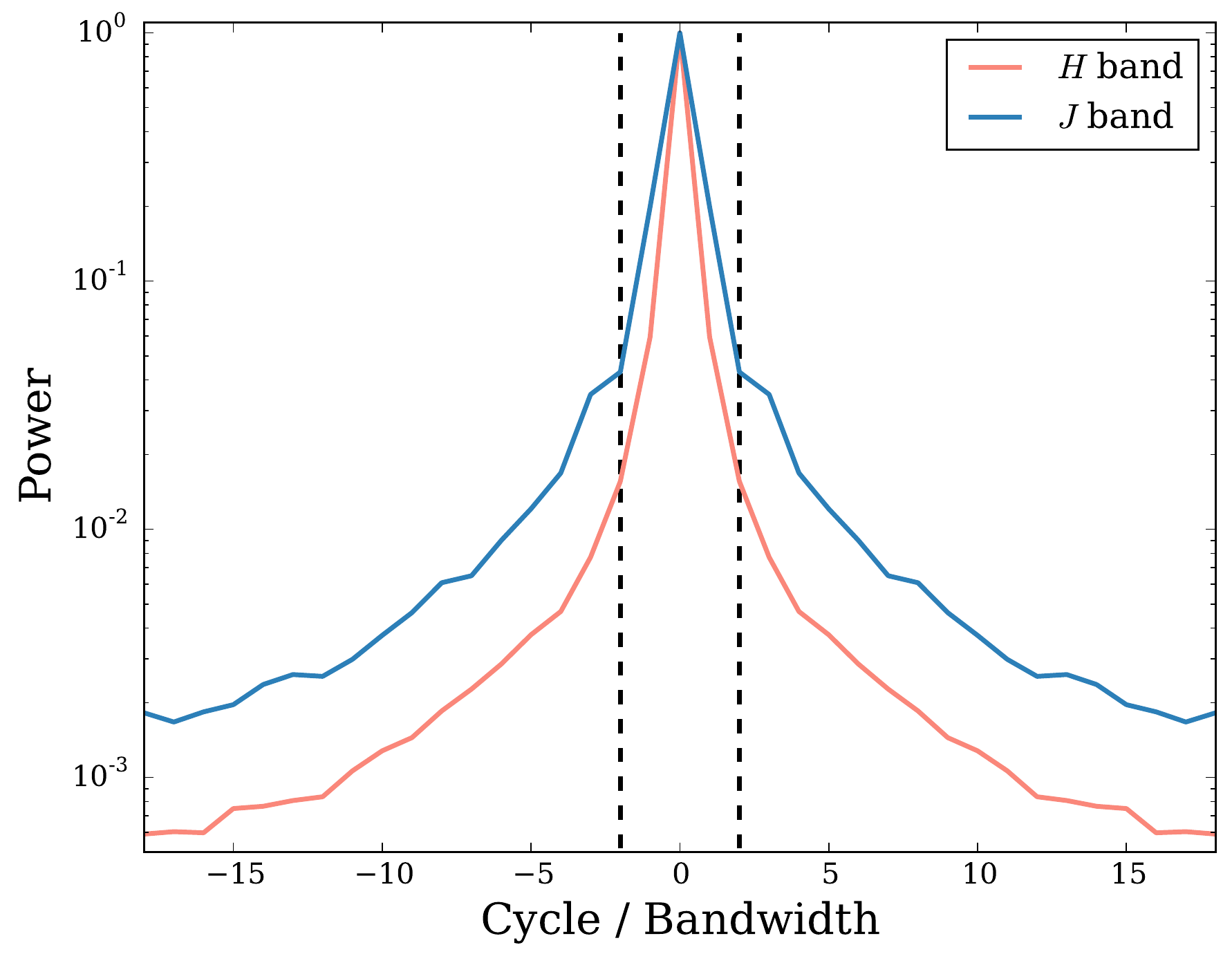}
\caption[Power Spectrum]{Power spectra as a function of cycles per bandwidth for each bandpass. The $H$ band has a bandwidth of 0.30 \micron~and $J$ band has a bandwidth of 0.23 \micron.  Vertical dashed lines indicate the boundary between low (between dashed lines) and high (outside dashed lines) frequency spectra.}
\label{pspec}
\end{figure}

\begin{figure}[h]
  \centering
  \begin{tabular}[b]{@{}p{.4\textwidth}@{}}
   \includegraphics[width=1.0\linewidth]{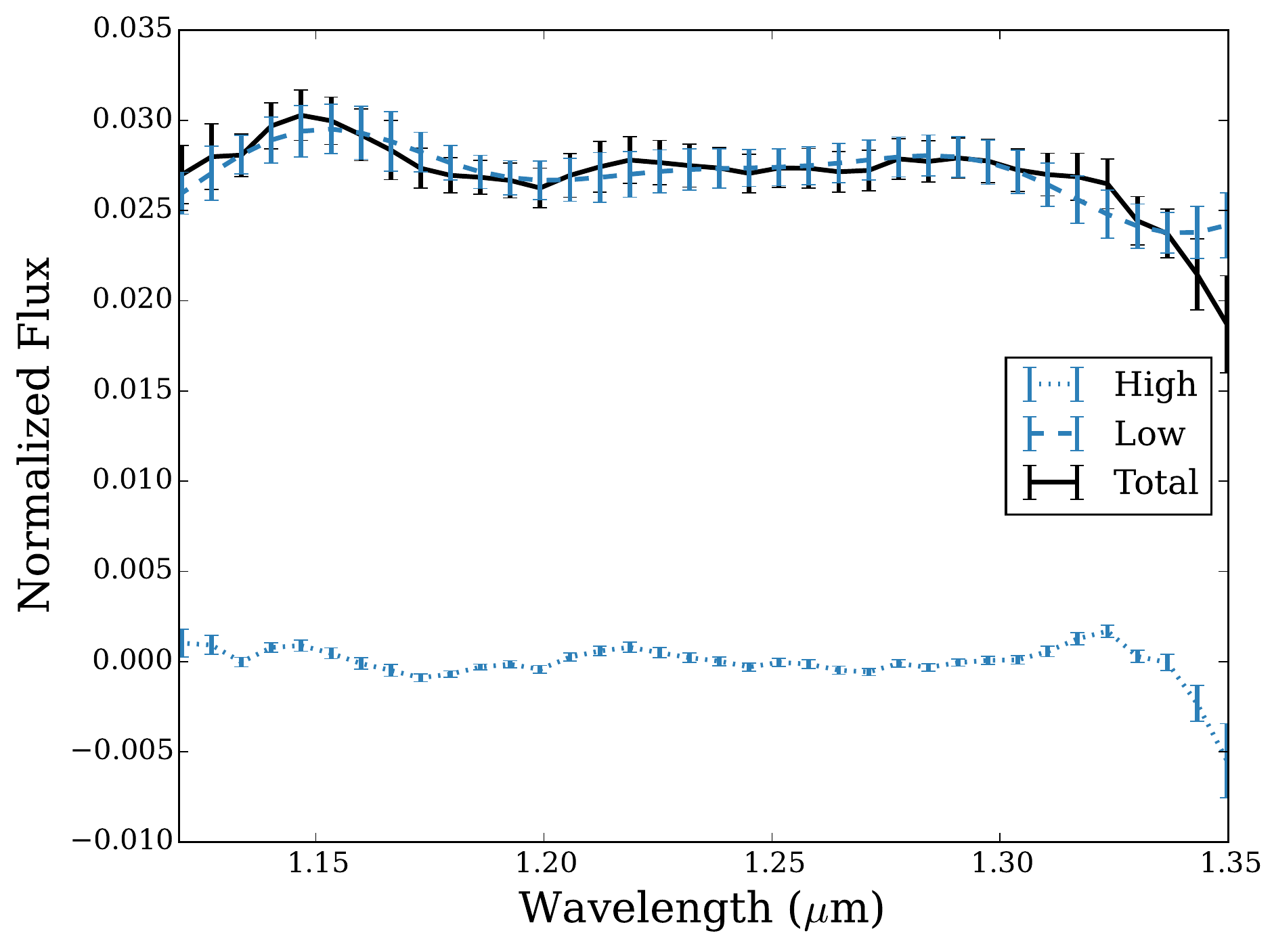} \\
    \centering\small (a)  
  \end{tabular}%
  \quad
  \begin{tabular}[b]{@{}p{.4\textwidth}@{}}
    \includegraphics[width=1.0\linewidth]{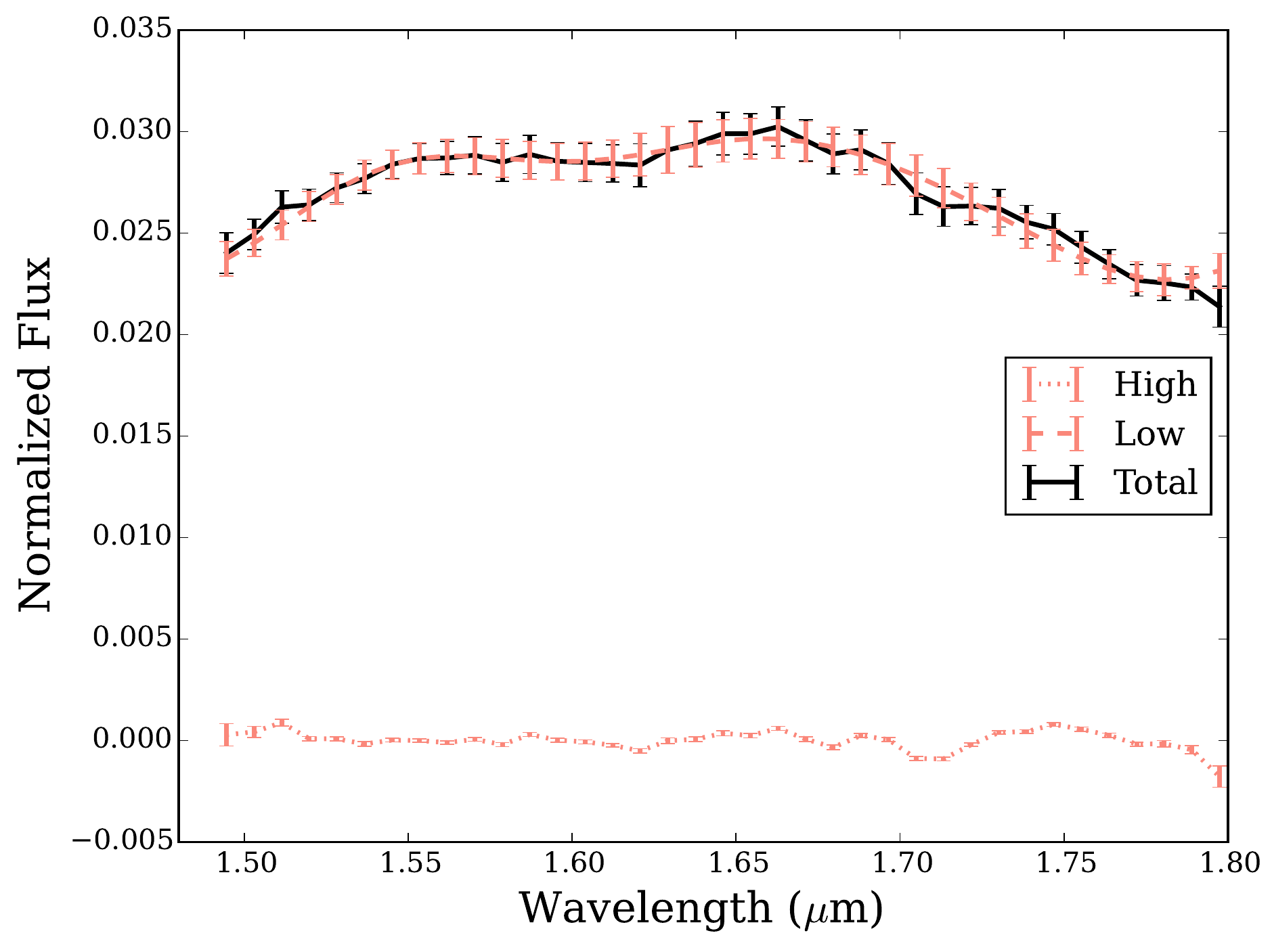} \\
    \centering\small (b) 
  \end{tabular} \\
  \caption[Spectral Splitting]{$J$ band (a) $H$ band (b) spectra split between low (dashed line) and high (solid line) frequencies.  Solid dark lines are the original spectra.}
  \label{splitspec}
\end{figure}

\begin{figure}[h]
\epsscale{.95}
\plotone{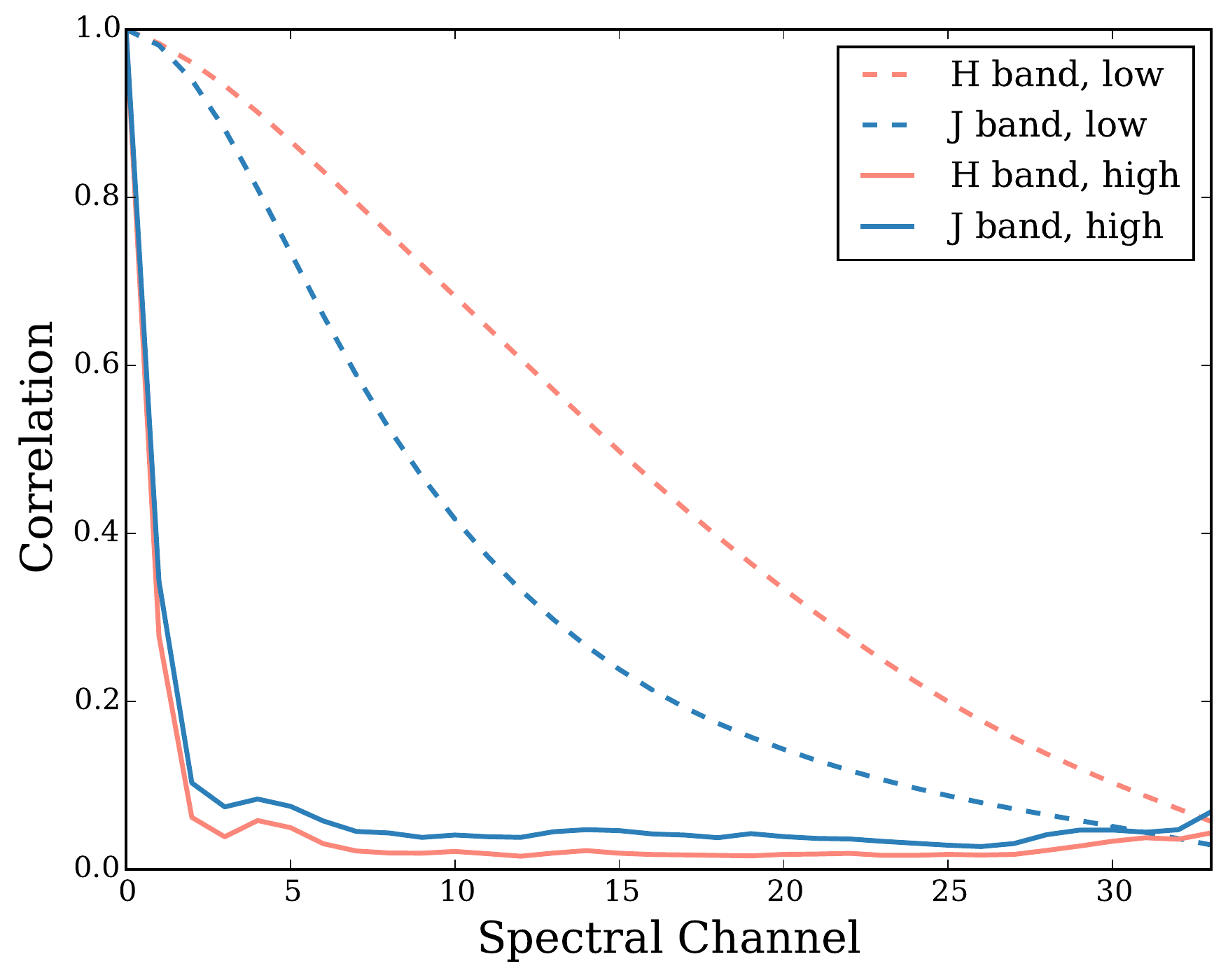}
\caption[Spectral Correlation]{Spectral correlation of the image noise as a function of wavelength channels for each filter.  The low frequency spectra (dashed lines) show a much higher covariance than the high frequency spectra (solid lines).}
\label{correlation}
\end{figure}

The $J$ and $H$ band spectra of HD~984~B were split into low and high frequencies by taking the Fourier transform of the spectra (see Figure~\ref{pspec}).  Any frequencies between -2 and 2 cycles per bandwidth are considered low frequencies, and anything outside of that range is designated as high frequencies.  This range was selected to incorporate the bulk of the flux in the low frequency range.  The high and low spatial frequency errors were propagated by taking the spectra at nine different position angles, splitting their high and low components and taking the spread in the noise of each component as the error.  The split high and low frequency spectra are shown in Figure~\ref{splitspec}. Testing the spectral noise correlation for the high frequency spectra separately found correlations covering just three wavelength channels for both bands as expected from read and background noises (see Figure~\ref{correlation}); this spectral correlation is consistent with the GPI pipeline spectral over-sampling.  The low frequency spectra were still highly correlated to fifteen and eight wavelength channels for $J$ and $H$ bands respectively (correlation reaching 50\%).

Since the spectral channels at different wavelengths are correlated, each spectrum can be binned accordingly to avoid biasing comparisons to the spectra of other objects.  High frequency spectra were binned by averaging three adjacent channels into twelve groups, with the last group averaging four channels to incorporate the leftover channel.  $H$ band low frequency spectra were binned into two groups of twelve channels each and one group of thirteen channels, and $J$ band low frequency spectra were binned into five groups, three of seven channels each and two of eight channels each.

The extracted GPI spectra of HD~984~B, and its low and high-frequency components, and the SINFONI {\it K} band spectrum from \citet{meshkat} were compared to a library of 1600 M, L, and T dwarf near-infrared spectra compiled from the SpeX Prism Library \citep{burgasser14}, the IRTF Spectral Library \citet{cushing}, the Montreal Spectral Library (e.g., \citealp{gagne,robert2016}), and the sample of young ultracool dwarfs presented in \citet{allers}. The spectrum of each object was convolved with a Gaussian to degrade it to the same resolution as the GPI data. For the comparison with the low and high-frequency components, the library spectra were processed through the same Fourier filtering and binning procedure as for the HD~984~B spectrum. For the fits to the individual bands, the scaling factor between the HD~984~B spectrum and each comparison object was found analytically by evaluating the derivative of the $\chi^2$ equation \citep[e.g.,][]{burgasser16}. The SINFONI {\it K} band spectrum was fit using a similar procedure, with the spectrum of HD~984~B and the comparison objects degraded to a resolution of $R\sim120$, similar to that of the majority of the objects within the spectral library. The SINFONI {\it K} band spectrum was not processed through the same Fourier filtering and binning steps as the GPI data.

The best fit to the three combined {\it JHK} spectra -- one each with the unfiltered {\it JH} spectrum, the low frequency {\it JH} spectrum ({\it JH$_{\rm low}$}), and the high frequency {\it JH} spectrum ({\it JH$_{\rm high}$}) -- was found by minimizing the goodness of fit statistic $G$ \citep{cushing08}, similar to the $\chi^2$ statistic but with each term of the summation weighted proportionally to the bandwidth of each channel within the spectrum (e.g., \citealp{burgasser10,gagne15}). This weighting was particularly important for the fit to the {\it JH$_{\rm low}$} spectrum which consisted of eight spectral channels, compared to the 160 within the {\it K} band spectrum. For each fit the same {\it K} band spectrum was used. Each band was fit with an independent scale factor, which accounts for uncertainties in the absolute flux calibration of the GPI data \citep{maire}, the differences in flux calibration between the GPI and SINFONI data, and for the spread in near-infrared colors seen for young brown dwarfs (Cruz et al. {\it submitted}).

\begin{figure*}
\epsscale{.8}
\plotone{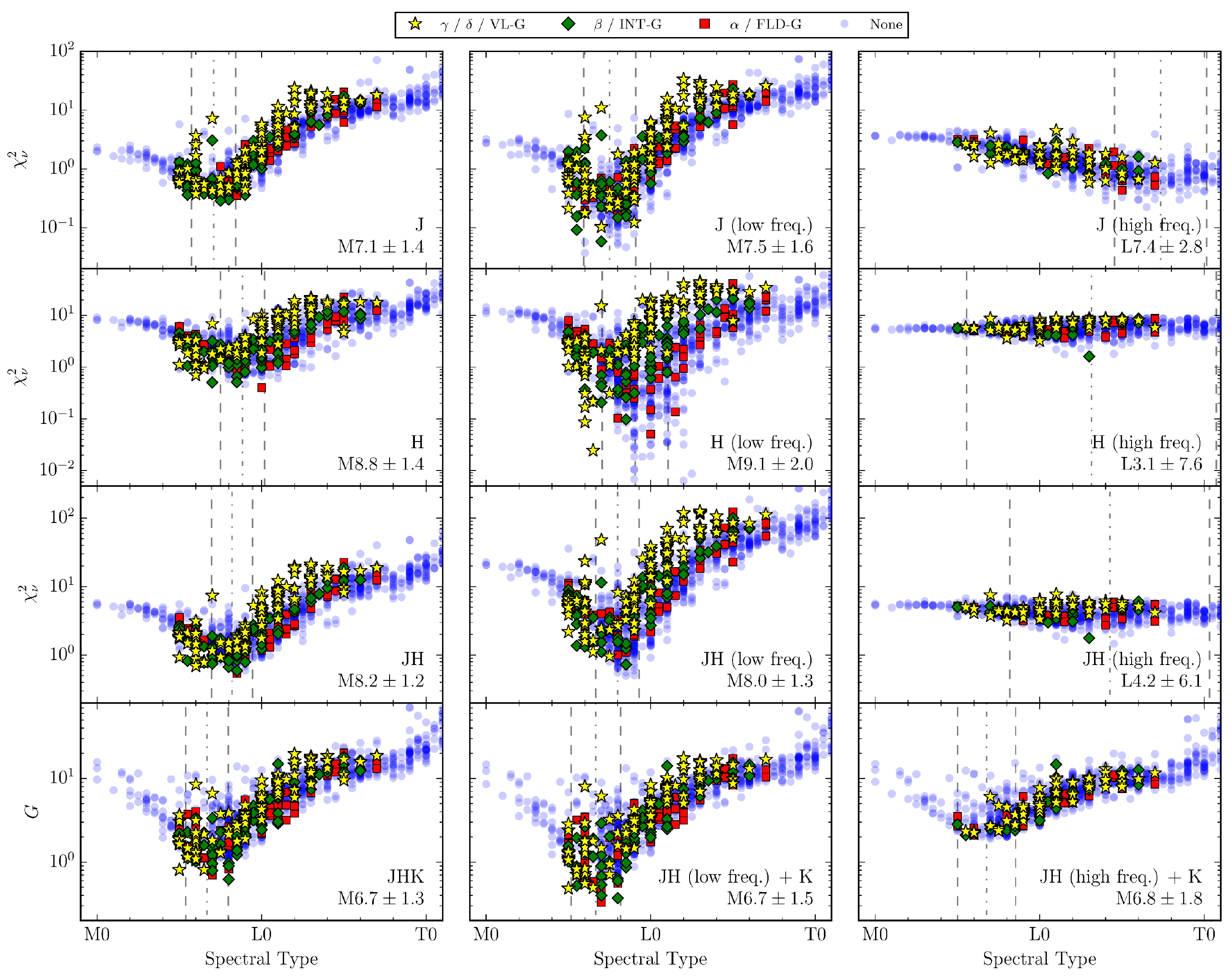}
\caption[$\chi^2$ Spectral Type Matching]{Goodness of fit statistic for each object within the spectral library using the spectrum of HD~984~B before filtering (left column), the low frequency component (middle column), and the high frequency component (right column). The top three rows show the reduced $\chi^2$ for the fit to the {\it J}, {\it H}, and {\it K} bands individually, while the bottom row shows the weighted goodness of fit statistic \citep[$G$,][]{cushing08} based on a fit to the full spectrum. The symbols denote the surface gravity classification of the comparison object: very low gravity ($\gamma$/$\delta$/{\sc vl-g}; yellow star), intermediate gravity ($\beta$/{\sc int-g}; green diamond), field gravity ($\alpha$/{\sc fld-g}; red square). Objects without a surface gravity classification are plotted as semi-transparent blue circles. The vertical dot-dashed and dashed lines denote the mean and the uncertainty of the spectral type of HD~984~B calculated for each fit.}
\label{chi}
\end{figure*}

The goodness of fit statistic for each comparison object is plotted as a function of spectral type for each fit for the unfiltered spectrum and the low and high frequency components in Figure~\ref{chi}. The spectral type of HD~984~B was estimated by calculating the goodness of fit statistic (either $\chi^2$ or $G$) of the spectrum of HD~984~B and a template spectrum from the library created by averaging all objects of a given spectral type. The weighted average of the spectral type of each template was adopted as the spectral type, with the individual weights drawn from the F-test probability distribution function \citep[e.g.,][]{burgasser10}. A systematic uncertainty of 1 subclass was assumed for all objects within the spectral library, and incorporated into the uncertainty on the spectral types given in Figure~\ref{chi}. The spectral types estimated from the fit to both the {\it JHK} and the {\it JH$_{\rm low}$K} spectra are consistent, both M$6.5\pm1.5$ when rounded to the nearest half subtype, suggesting that the covariance between neighbouring wavelength channels within the GPI data did not strongly bias the fit. While the spectral type from the fit to the {\it JH$_{\rm high}$K} spectrum was also consistent at M$7\pm2$, this was almost entirely driven by the {\it K} band fit as neither the {\it J} nor {\it H} band GPI spectra of HD 984 exhibited strong spectral features, as demonstrated by the very large uncertainty on the spectral type from the {\it JH$_{\rm high}$} fit (L$4\pm6$). While the high frequency spectrum did not help constrain the spectral type of HD~984~B, a similar analysis may prove more useful for {\it K} band spectra which cover the CO band heads at 2.29$\mu$m and 2.32$\mu$m \citep[e.g.,][]{konopacky13}.

\begin{figure}[h]
  \centering
  \begin{tabular}[b]{@{}p{0.45\textwidth}@{}}
   \includegraphics[width=1.0\linewidth]{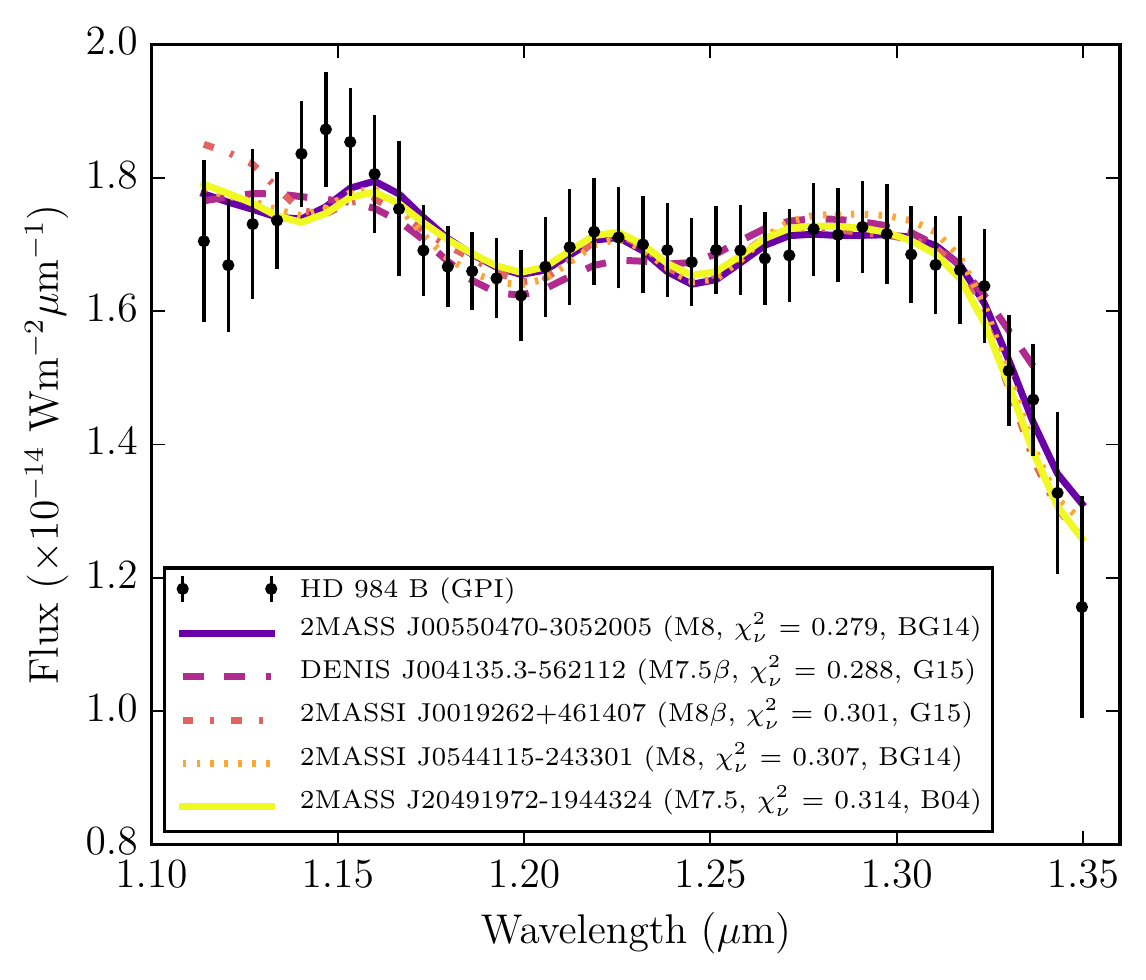} \\
  \end{tabular}%
  \quad
  \begin{tabular}[b]{@{}p{0.45\textwidth}@{}}
    \includegraphics[width=1.0\linewidth]{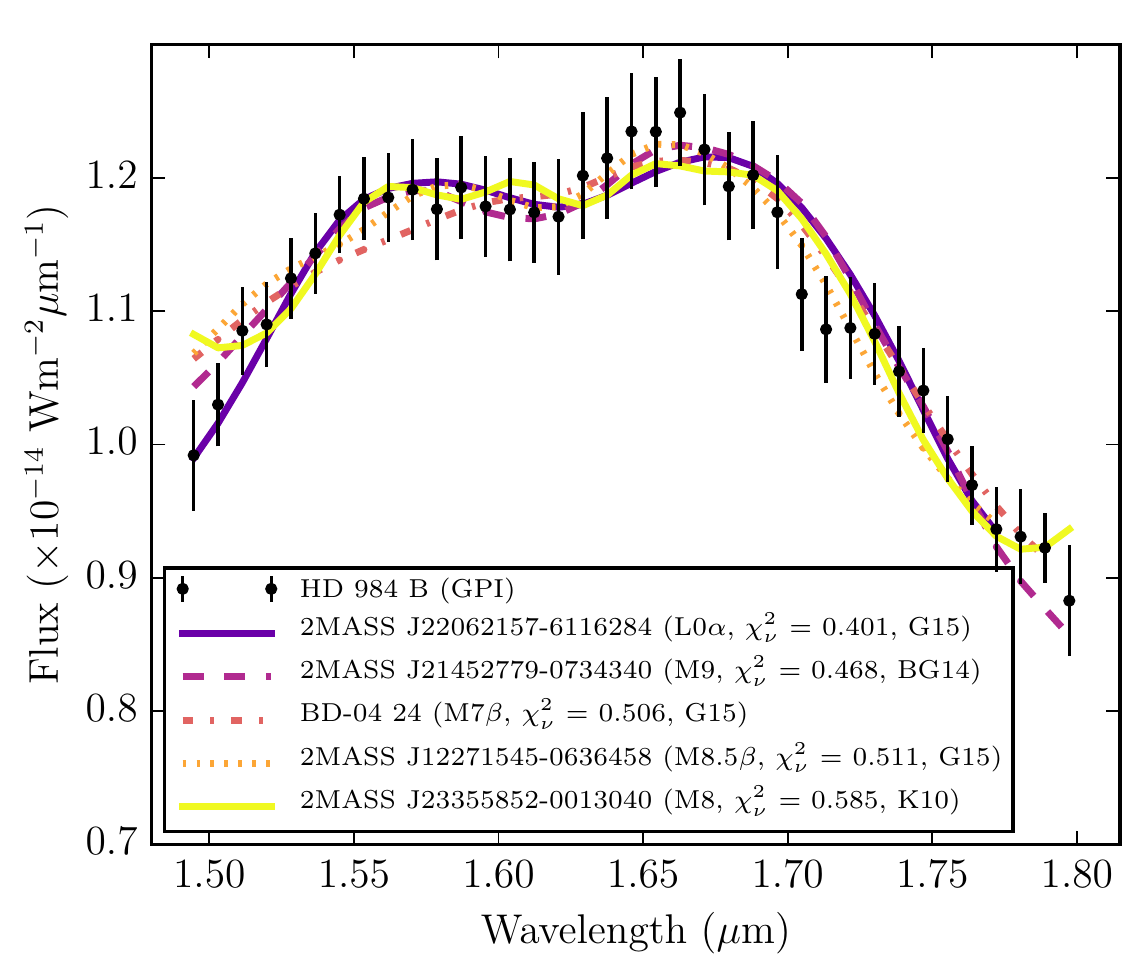} \\
  \end{tabular} \\
  \caption{GPI $J$ band (left panel) and $H$ band (right) spectra of HD~984~B (black points) plotted with the spectra of the five best fit objects found for each band degraded to the resolution of GPI. The name, spectral type, and reference for the spectra are given in the legend (B04: \citealp{burgasser}, BG14: \citealp{bardalez}, G15: \citealp{gagne15}, K10: \citealp{kirkpatrick}). Considering only the $J$ band the spectral type of the five best fit objects ranged between M7.5 and M8, with two having indications of a surface gravity lower than field objects. For the $H$ band, the best fit objects span a larger range of spectral types, M7 to L0, with two having intermediate surface gravity classification }
  \label{hjspec}
\end{figure}

\begin{figure*}
\epsscale{1.1}
\plotone{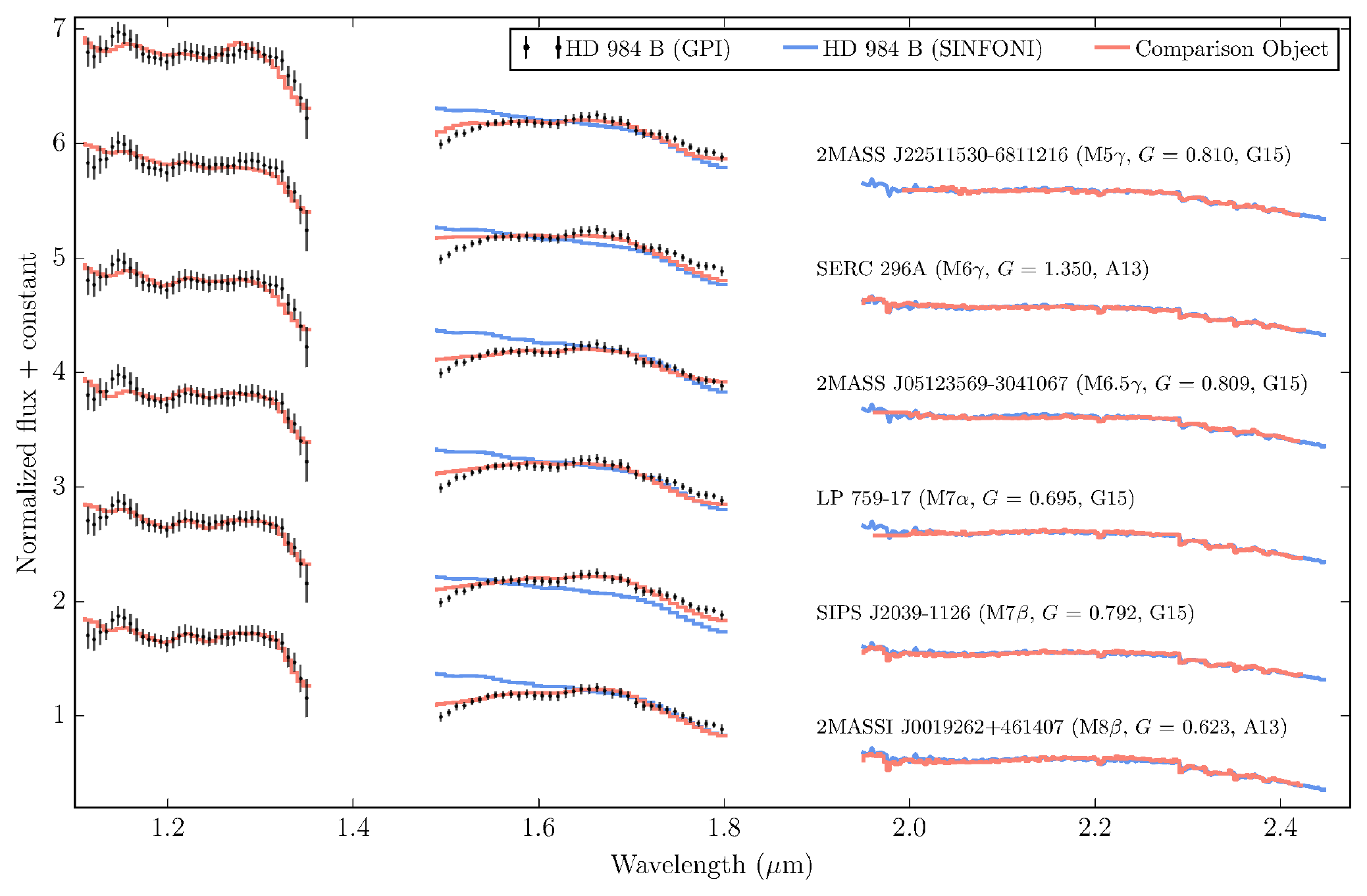}
\caption{GPI {\it JH} (black points) and SINFONI {\it HK} (blue curve; \citealp{meshkat}) spectra of HD~984~B compared with the five objects best fit to the unfiltered {\it JHK} spectrum ordered by ascending spectral type. Also plotted is the spectrum of SERC 296A, the best fit to the SINFONI {\it HK} spectrum identified by \citet{meshkat}. The {\it JHK} spectrum of each of the six comparison objects was multiplied by the scale factor which minimized the $\chi^2$ between the comparison object and the GPI spectrum of HD~984~B at {\it H} band. As the individual bands were allowed to float independently during the fitting procedure, the GPI {\it J} band and SINFONI {\it K} spectra of HD~984~B were scaled to minimize the $\chi^2$ within each band. The SINFONI {\it H} band spectrum was scaled by the same factor as the {\it K} band.  The cause of the difference between the GPI and SINFONI {\it H} band spectra at the shortest wavelengths is not yet known.  No additional scaling was performed on any band of the comparison object spectra. Comparison spectra were obtained from a number of sources (A13: \citealp{allers}, G15: \citealp{gagne15}).}
\label{spec}
\end{figure*}

The five objects within the spectral library which best fit the unfiltered {\it J} and {\it H} band spectra of HD~984~B are shown in Figure~\ref{hjspec}, and are consistent with the spectral type estimate from each band given in Figure~\ref{chi}. For the {\it J} band fit, two of the objects are known members of young moving groups: \objectname{DENIS J004135.3-562112} (M7.5$\beta$, Tucana--Horologium 20--40~Myr; \citealp{reiners09}) and \objectname{2MASSI J0019262+461407} (M8$\beta$, AB Doradus 30--50~Myr; \citealp{schlieder12}). Additionally, \objectname{2MASS J20491972-1944324} was listed by \citet{gagne14} as having indications of youth based on both a red color and fits to model atmospheres. For the {\it H} band fit, the best fit object has a no indication of low surface gravity \citep{gagne}, and of the five best fit only one is a candidate member of a young moving group, \objectname{2MASS J12271545-0636458} (M8.5$\beta$, TWA 5--15~Myr; \citealp{gagne}). The five best fit objects to the {\it JHK} spectrum are shown in Figure~\ref{spec}. The best fit object was \objectname{2MASS J0019262+461407} (M$8\beta$), a likely member of the AB Doradus moving group \citep{schlieder12}, consistent with the $115\pm85$~Myr age estimate for HD~984 \citep{meshkat}. Of the remaining four, two are classified as having low surface gravity and are likely members of the Tucana--Horologium (\objectname{2MASS J225511530-6811216}, M$5\gamma$) and Columba (\objectname{2MASS J05123569-3041067}, M6.5$\gamma$) moving groups \citep{gagne}, one has intermediate surface gravity and is a likely member of the AB Doradus moving group (\objectname{SIPS J2039-1126}, M$7\beta$; \citealp{gagne}), and one has a field surface gravity classification (\objectname{LP 759-17}, M$7\alpha$; \citealp{gagne}).


Based on the spectral fitting alone there is not a conclusive result in terms of the surface gravity classification \citep[e.g.][]{cruz09} for HD~984~B. While the global minima of $G$ for both the {\it JHK} and {\it JH$_{\rm low}$K} spectra are populated by field or intermediate gravity objects, the low gravity objects appear to form a sequence displaced from field gravity objects by approximately one to two spectral subtypes hotter. Performing the same spectral typing procedure outlined previously on the {\it JHK} spectrum of HD~984~B using the subset of field gravity objects (gravity classification of $\alpha$ or {\sc FLD-G}) yields a spectral type of M$7\pm1$, compared with M$6\pm1$ when using the low gravity subset ($\gamma$/$\delta$/{\sc vl-g}). This effect is most pronounced for the {\it H} band where surface gravity has a significant effect on spectral morphology, with the same analysis yielding a spectral type of M$8.5\pm1.5$ using field gravity objects, and of M$6.5\pm1.0$ using low gravity objects. As the difference between the minimum goodness of fit statistic of the two subsets is small ($G=0.70$ and $G=0.81$ respectively), it is not possible to assign either a field or low surface gravity classification to HD~984~B based on this analysis.

Photometric and other spectral indicators can also be used to classify the gravity of late M to early L type objects \citep{allers,filippazzo}.  We test the H-cont index described in \citet{allers} which uses the blue end slope of the H band continuum.  A higher index ($\sim1.0$) indicates a straighter slope which is seen in lower gravity objects.  HD 984 B, when compared to other objects, has a comparable H-cont index ($0.95\pm0.01$) for its determined spectral type, and there is no indication it is of low gravity.  Furthermore, comparisons of HD 984 B's $J$ band magnitude versus spectral type \citep[see][]{filippazzo} is consistent with objects from the BDNYC catalogue (\url{doi:10.5281/zenodo.45169}) and does not show low gravity signatures.  Similarly, a CMD of $J$ vs $J-K$ magnitudes shows HD 984 B at a similar location as other late M types, instead of shifted redder which would indicate low gravity. From these indicators, there is no conclusive evidence showing HD 985 B has low surface gravity which suggests that the true age for the system may be at the high end of the age estimate of $115\pm85$ Myr \citep{meshkat}.

We note a significant difference between the GPI and SINFONI spectra towards the blue end of the $H$ band. The cause for this difference is unknown but may arise from differences in data processing and speckle subtraction artifacts. A complete analysis of these differences is beyond the scope of this paper. 

\begin{figure}
\epsscale{1.1}
\plotone{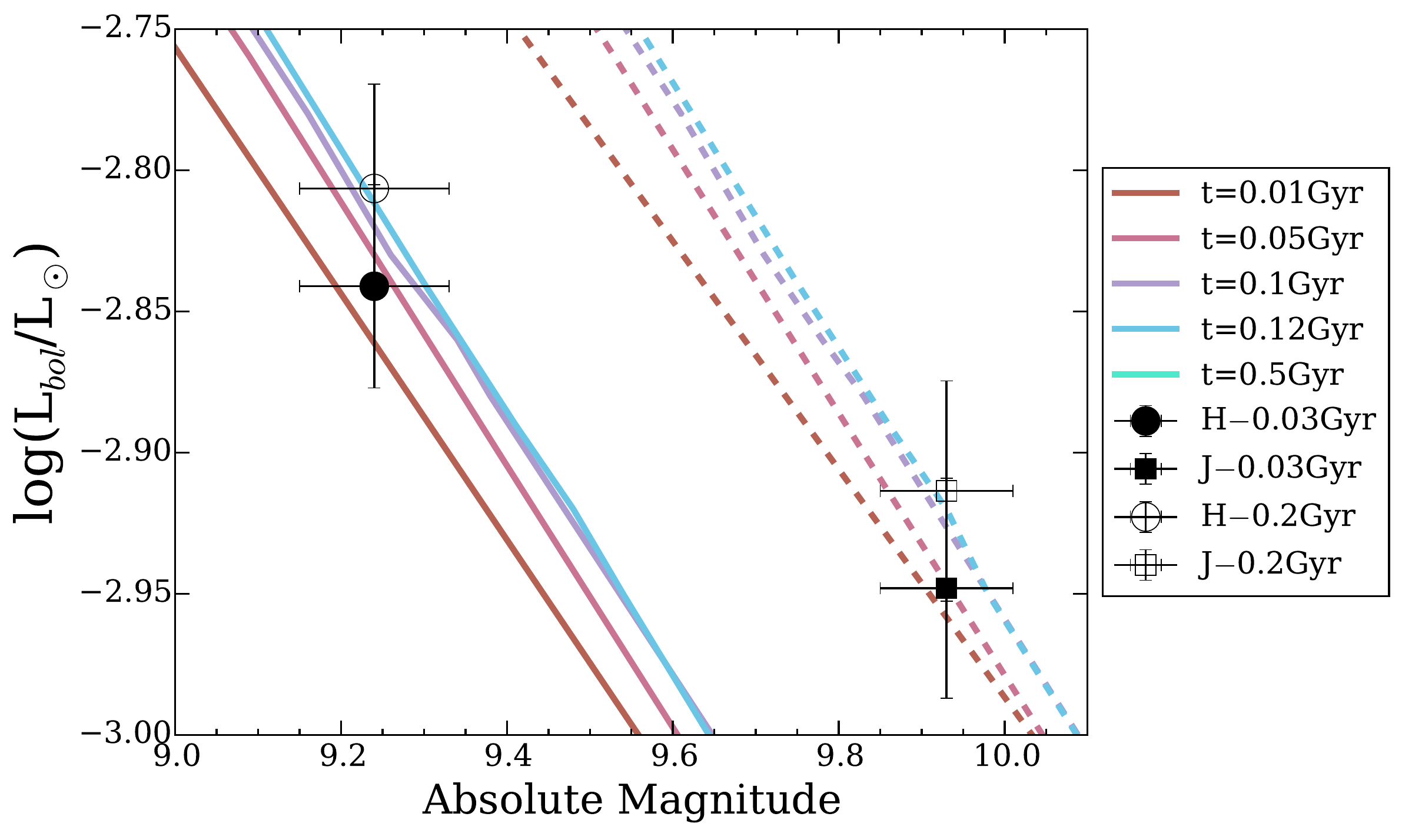}
\caption[Luminosity Models]{DUSTY luminosity models for 0.01 Gyr to 0.5 Gyr plotted with HD 984 B data. The luminosities for $J$ and $H$ band data were derived through interpolating the models at 30 and 200 Myr. Isochrone models shown were selected to encompass the age range of HD 984 B and are colour coded by age with blue being older models and red younger models. $J$ band models are shown by dashed lines and $H$ by solid. HD 984 B is show by black points for an age of 0.03 Gyr and open points for 0.2 Gyr.  Circles represent the H band and squares the J band. 
}
\label{lummodel}
\end{figure}

\begin{figure}
  \centering
  \begin{tabular}[b]{@{}p{0.35\textwidth}@{}}
   \includegraphics[width=1.0\linewidth]{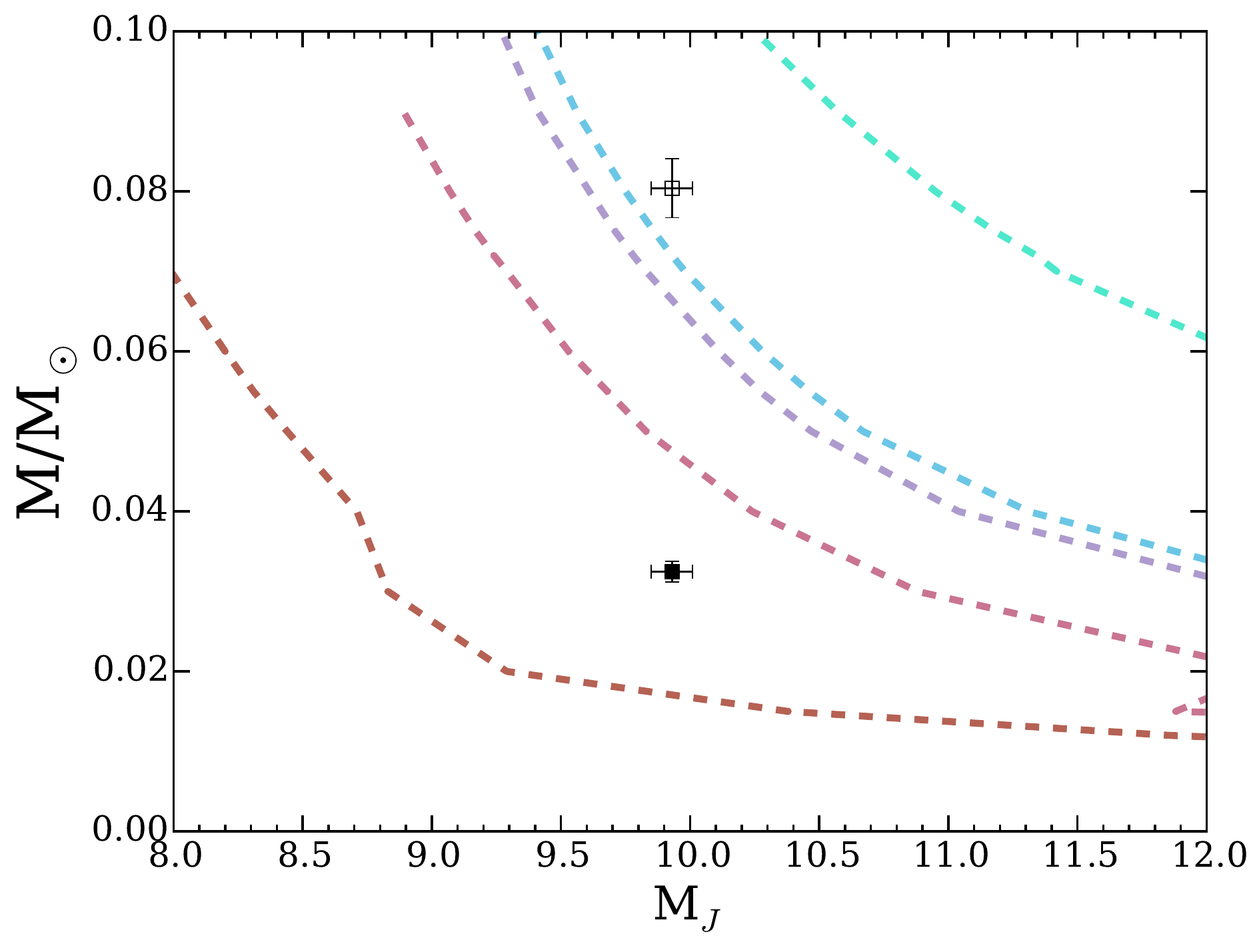} \\
    \centering\small (a)  
  \end{tabular}%
  \quad
  \begin{tabular}[b]{@{}p{0.35\textwidth}@{}}
    \includegraphics[width=1.0\linewidth]{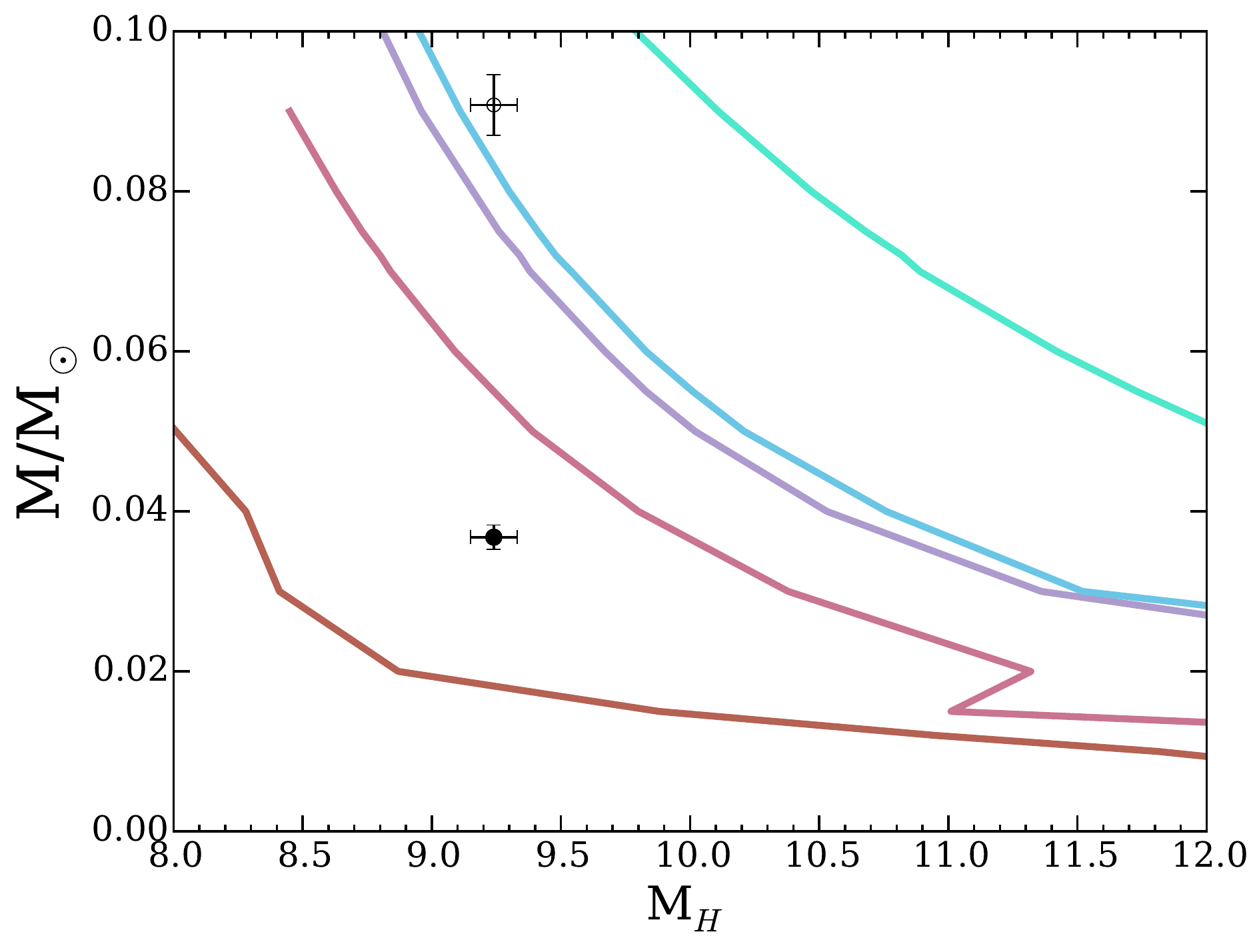} \\
    \centering\small (b) 
  \end{tabular} \\
  \caption[Mass Models]{DUSTY mass models for $J$ band (a) and $H$ band (b) along with the interpolated mass of HD 984 B at 30 Myr and 200 Myr. Colours and symbols the same as Figure~\ref{lummodel}. 
  }
  \label{massmodel}
\end{figure}

In addition to spectral type matching, the photometry can also be analyzed to determine the object's luminosity, temperature and mass.  These calculations from the new GPI data can be used both to compare to the earlier SINFONI estimates of luminosity, temperature and mass, and to provide better estimates with the additional bands. Since \citet{meshkat} find no significant difference between evolutionary models, we use DUSTY isochrone models \citep{chabrier}. As these models are highly dependent on age since low-mass objects slowly cool with time, it is necessary to have an accurate age estimate. Using the detailed analysis from \citet{meshkat}, who made a comprehensive age estimate, we adopt the same age range, 30--200 Myr, for HD 984 B.  Taking the DUSTY isochrone models, which provide tables of the luminosity and magnitudes for a range of planet masses at particular ages, we first interpolate between model ages to generate new finer grid tables for 10 to 500Myr. This is then used to derive the predicted luminosities at 30 and 200 Myr that correspond to the measured values of J and H. Luminosity and mass uncertainties are propagated from uncertainties in the absolute magnitudes.  Although the age of the system is inconsequential when computing luminosity (see Figure~\ref{lummodel}), it is highly influential when estimating mass (see Figure~\ref{massmodel}).  The luminosity, accounting for the age range and both bands is $\log(L_{\mathrm{bol}}/L_{\sun})=-2.88\pm0.07$ dex, in agreement with \citet{meshkat}. 
  We use the same technique to find the mass, and we find the $H$ band absolute magnitude corresponds to a range of masses from $39\pm2$ M$_{\mathrm{Jup}}$ at 30 Myr to $94\pm4$ M$_{\mathrm{Jup}}$ at 200 Myr. The $J$ band yields masses of $34\pm1$ M$_{\mathrm{Jup}}$ and $84\pm4$ M$_{\mathrm{Jup}}$ for the same ages.  Temperature analysis, conducted in the same manner as the luminosity models and using the same DUSTY models, found object temperatures of $2458\pm32$~K to $2800\pm37$~K for $J$ band over the same age range.  The $H$ band magnitude give temperatures of $2545\pm28$~K to $2896\pm31$~K. The low photometric uncertainties correspond to the low quoted temperature errors, but much larger systematic errors likely are present due to model uncertainties.  Using the spectral type estimated previously (M$6.5\pm1.5$) and the spectral type-to-temperature conversion from \citet{stephens}, we derive an effective temperature of $T_{\rm eff}=2730^{+120}_{-180}$ K, consistent with the results of \citet{meshkat}, and intermediate to the temperature estimates from the {\it J} and {\it H} band absolute magnitudes. 
  
Due to extensive spectral covariance, more rigorous characterization estimation through spectroscopy is unfeasible. Instead, we further empirically compare HD 984 B to other objects fitted by colour, which are independent of models. For this, we use field objects from \citet{filippazzo}.  Using HD 984 B's best matched specturm type, M6.5, we estimate $L_{\mathrm{bol}}=-3.1^{+0.1}_{-0.2}$.



\begin{deluxetable*}{lccccc}
\tabletypesize{\scriptsize}
\tablewidth{0pt}
\setlength{\tabcolsep}{3pt}
\tablecaption{System Properties}
\tablehead{{Property} & {unit} & {HD 984} &{HD 984 B$^a$}& {HD 984 B$^b$}&{References}}
\startdata
\hline
Distance & pc & $47.1\pm1.4$ &&&1\\
Age &Myr & 30--200&&&2\\
m$_\mathrm{H}$ &-- &  $6.170\pm0.038$   & $12.58\pm0.05$ &$12.60\pm0.05$&3\\
m$_\mathrm{J}$& -- & $6.402\pm0.023$ & &$13.28\pm0.06$&3\\
Spectral Type & -- &  F7V &  M$6.0\pm0.5$ &M$6.5\pm1.5$ &4\\
Temperature & K & $6315\pm89$  & $2777^{+127}_{-130}$&$2730^{+120}_{-180}$&5 \\
$\log(L_{\mathrm{bol}}/L_{\sun})$ & dex &$0.346\pm0.027$ & $-2.815\pm0.024$&$-2.88\pm0.07$&2 \\
Mass& -- & $\sim$1.2$M_\sun$ & $33\pm6$ to $94\pm10$ M$_{\mathrm{Jup}}$ &$34\pm1$ to $94\pm4$ M$_{\mathrm{Jup}}$  &2\\
Semi Major Axis$^c$ & au &&& $18_{-4}^{+10}$\\
Period$^c$ & yrs  &&&  $70_{-25}^{+69}$ \\
Inclination$^c$ & deg &&& $119_{-5}^{+6}$\\
Eccentricity$^c$& --  &&&$0.18_{-0.13}^{+0.29}$\\

\enddata
\label{props}
\tablecomments{$^a$Results from \citet{meshkat}; $^b$New results presented in this paper; $^c$Ranges listed encapsulate the 68\% confidence interval. References for primary's properties from: (1) \citet{van}, (2) \citet{meshkat}, (3) \citep{cutri}, (4) \citet{houk}, (5) \citet{casagrande}}
\end{deluxetable*}

\section{Conclusion}\label{conc}

Our new observations of HD~984~B with the Gemini Planet Imager have built upon the results presented by \citet{meshkat} and provided a more comprehensive understanding of this young substellar companion.  Here we summarize the characterization we have achieved in our analysis.  All numerical results can be found in Table~\ref{props}.

With a three year baseline between the first epoch of \citet{meshkat} astrometry and the new GPI astrometry presented in this work, we derive the first constraints on all of the orbital parameters for this system using a rejection sampling technique.  Continued astrometric monitoring of HD~984~B with GPI will help to further constrain these orbital parameters, and reduce the effects of any systematic biases between measurements obtained from different instruments.

The GPI observations were also used to investigate the photometric and spectroscopic properties of HD~984~B, which were compared to predictions of evolutionary models and to other substellar objects spanning a range of spectral types and ages. From the integrated {\it J} and {\it H} band spectra, and knowing the distance to the object, we measure an absolute magnitude which is used with DUSTY evolutionary models to derive a luminosity, mass and temperature. 

Complementing our GPI {\it JH} spectrum with the SINFONI {\it K} band spectrum presented in \citet{meshkat}, we estimated the spectral type of HD~984~B from spectral templates constructed from a large number of near-infrared spectra of low-mass stars and brown dwarfs. To account for spectral covariances, the spectra were split into high and low spatial frequencies and binned according to the correlation length measured within each spectrum.  The best spectrum match was further used to calculate an effective temperature using empirical spectral type-to-temperature relations.  While the results from the fit to the unfiltered and low frequency spectra were consistent in terms of the derived spectral type, the high frequency component did not help constrain the spectral type due to the lack of sharp spectral features. This method however, may prove useful to match spectral features with $K$ band data, where narrow spectral features, such as CO, can be identified and fitted.  Splitting the spectra in these cases will allow for better noise statistics and improved $\chi ^2$ analysis.

The surface gravity of HD~984~B was also investigated, as a low surface gravity could have been used to provide a further constraint on the age and mass of the system. The five best fitting objects to the {\it JHK} spectrum of HD~984~B had a mix of gravity classifications. With this ambiguous result, several photometric and spectral indicators were also computed to look for additional evidence of low surface gravity. None of these indicators suggested a low surface gravity for HD~984~B, and we therefore do not assign a gravity classification to the spectral type.

HD~984~B is one of the latest of a growing number of brown dwarf and low-mass companions discovered via direct imaging, often serendipitously during searches for exoplanets. As demonstrated for this object, substellar companions discovered in these campaigns can be rapidly characterized with an integral field spectrograph, and preliminary constraints on orbital parameters can be derived with a relatively short baseline between epochs. With the many ongoing surveys using extreme adaptive optics instruments such as GPI, SPHERE, and SCExAO, there is ample opportunity to discover and characterize new substellar objects in the near future. Although one of object alone cannot prove a rule, continued identification and characterization of this class of objects will undoubtedly further our understanding of their formation.

\acknowledgments
\section*{Acknowledgments}
We would like to thank Dr.~Tiffany Meshkat for providing the SINFONI $K$ band spectra. This research has benefited from the SpeX Prism Library and SpeX Prism Library Analysis Toolkit, maintained by Adam Burgasser at \url{http://www.browndwarfs.org/spexprism},  the Montreal Brown Dwarf and Exoplanet Spectral Library, maintained by Jonathan Gagn\'e, and the SIMBAD database, operated at CDS, Strasbourg, France.  This paper is based on observations obtained at the Gemini Observatory, which is operated by the Association of Universities for Research in Astronomy, Inc., under a cooperative agreement with the NSF on behalf of the Gemini partnership: the National Science Foundation (United States), the National Research Council (Canada), CONICYT (Chile), Ministerio de Ciencia, Tecnolog\'{i}a e Innovaci\'{o}n Productiva (Argentina), and Minist\'{e}rio da Ci\^{e}ncia, Tecnologia e Inova\c{c}\~{a}o (Brazil). This publication additionally makes use of data products from the Two Micron All Sky Survey, which is a joint project of the University of Massachusetts and the Infrared Processing and Analysis Center/California Institute of Technology, funded by the National Aeronautics and Space Administration and the National Science Foundation.  Portions of this work were performed under the auspices of the U.S. Department of Energy by Lawrence Livermore National Laboratory under Contract DE-AC52-07NA27344. Supported by NASA grants NNX11AD21G and NNX15AD95/NEXSS (R.J.D.R., J.R.G, J.J.W, T.M.E, P.G.K.) and NX14AJ80G (E.L.N., S.B, F.M.), and NSF grants AST-0909188 and AST-1313718 (R.J.D.R., J.R.G., J.J.W., T.M.E., P.G.K.)


\bibliographystyle{plainnat}

\end{document}